\documentclass[preprintnumbers,nofootinbib,floatfix]{revtex4-1}

\usepackage{graphicx}
\usepackage{hyperref}
\newcommand{\gsim}{\gtrsim}

\begin{document}

% page numbers bottom-center
\pagestyle{plain}

%%%%%%%%%%%%%%%%%%%%%%%%%%%%%%%%%%%%%%%%%%%%%%%%%%%%%%%%%%%%%%%%%%%%%%%%%%%%

\title{On the determination of anti-neutrino spectra from nuclear reactors}

\author{Patrick Huber
\footnote{email: pahuber@vt.edu}
}

\affiliation{Center for Neutrino Physics, Department of Physics,
  Virginia Tech, Blacksburg, VA 24061, USA}

%%%%%%%%%%%%%%%%%%%%%%%%%%%%%%%%%%%%%%%%%%%%%%%%%%%%%%%%%%%%%%%%%%%%%%%%%%%%

\begin{abstract}
  In this paper we study the effect of, well-known, higher order
  corrections to the allowed beta decay spectrum on the determination
  of anti-neutrino spectra resulting from the decays of fission
  fragments. In particular, we try to estimate the associated theory
  errors and find that induced currents like weak magnetism may
  ultimately limit our ability to improve the current accuracy and
  under certain circumstance could even largely increase the
  theoretical errors. We also perform a critical evaluation of the
  errors associated with our method to extract the anti-neutrino
  spectrum using synthetic beta spectra.  It turns out, that a fit
  using only virtual beta branches with a judicious choice of the
  effective nuclear charge provides results with a minimal
  bias. We apply this method to actual data for $^{235}$U, $^{239}$Pu
  and $^{241}$Pu and confirm, within errors, recent results, which
  indicate a net 3\% upward shift in energy averaged anti-neutrino
  fluxes. However, we also find significant shape differences which
  can, in principle, be tested by high statistics anti-neutrino data
  samples. 
\end{abstract}
\maketitle

%%%%%%%%%%%%%%%%%%%%%%%%%%%%%%%%%%%%%%%%%%%%%%%%%%%%%%%%%%%%%%%%%%%%%%%%%%%

\section{Introduction}

Neutrino physics as a branch of experimental physics started with
Cowan's and Reines' detection of neutrinos produced in a nuclear
reactor~\cite{Cowan:1992xc}. Since then, reactor neutrino experiments
have played a  crucial role in shaping our understanding of the
physical properties of the neutrino\footnote{Throughout this paper we
  will refer to both neutrino and anti-neutrino simply as neutrino and rely on the context for disambiguation.},
for a review see {\it e.g.} Ref.~\cite{Bemporad:2001qy}, and will continue
to do so in the future~\cite{Ardellier:2006mn,Guo:2007ug,Ahn:2010vy}.
In a nuclear reactor there are about 6 $\beta$-decays, and hence
neutrinos, per fission or about $2\times10^{20}$ neutrinos per second
per GW of thermal power. Fortunately, there are only four isotopes
whose fission make up more than 99\% of all reactor neutrinos with an
energy above the inverse $\beta$-decay threshold: $^{235}$U,
$^{239}$Pu, $^{241}$Pu and $^{238}$U. Nonetheless, the resulting
neutrino flux is a superposition of thousands of $\beta$-decay
branches of the fission fragments of those four isotopes and thus, a
first principles calculation is challenging, even with modern nuclear
structure data~\cite{Mueller:2011nm}.  Therefore, reactor neutrino
fluxes from the thermal fission of $^{235}$U, $^{239}$Pu and
$^{241}$Pu are obtained by inverting measured total $\beta$-spectra
which have been obtained in the 1980s at the Institut Laue-Langevin
(ILL)~\cite{Schreckenbach:1985ep,VonFeilitzsch:1982jw,Hahn:1989zr}.
There are, currently, no corresponding total $\beta$-spectra for
$^{238}$U since this isotope is fissioned only by fast neutrons.
Modern multi-baseline
experiments~\cite{Ardellier:2006mn,Guo:2007ug,Ahn:2010vy} are designed
to be independent of a very precise flux knowledge at the source,
however many of the previous measurements employed only one
baseline~\cite{Bemporad:2001qy} and initial data from Double
Chooz~\cite{Ardellier:2006mn} and Daya Bay~\cite{Guo:2007ug} most
likely will be single baseline data, too.

Recently, a reevaluation of reactor neutrino spectra has been
performed~\cite{Mueller:2011nm} and a significant upward shift
in the predicted fluxes of about 3\% was found. This in turn translates into a
weakening of existing reactor bounds on the absence of $\bar\nu_e$
disappearance with wide ranging consequences on the possible existence
of a sterile neutrino with a $\Delta m^2\gsim
1\,\mathrm{eV}^2$~\cite{Mention:2011rk,Kopp:2011qd}. Here, we present
an independent inversion of the ILL $\beta$-spectra into neutrino
spectra for the isotopes $^{235}$U, $^{239}$Pu and $^{241}$Pu. To this
end, we first review various corrections to the allowed $\beta$-decay
shape and estimate the associated theory errors in
section~\ref{sec:wm}. We will use virtual $\beta$-branches and apply
the corrections obtained in section~\ref{sec:wm} individually for each
branch. We then use synthetic data sets and Monte Carlo simulations to
quantify the bias and statistical errors associated with the inversion
procedure in section~\ref{sec:bias}. In section~\ref{sec:zeff} we
compute the effective nuclear charge $\bar Z$ and provide error
estimates. We then apply the method which we have developed to the
actual $\beta$-data and present our result on the neutrino fluxes in
section~\ref{sec:results}. In the following
section~\ref{sec:discussion} we perform a critical comparison of our
result with previously obtained neutrino fluxes and point out a possible
explanation for the flux shift in terms of known nuclear physics.
Finally, we conclude in section~\ref{sec:conclusion}.

\section{Beta spectra}
\label{sec:beta}

In this section we describe corrections to the allowed
$\beta$-spectrum shape and provide an estimate of the associated
uncertainty. Here, we mainly follow the notation of
Ref.~\cite{Wilkinson:1988}. The allowed $\beta$-spectrum is given
by~\cite{Fermi:1934hr}
\begin{equation}
\label{eq:simplebeta}
N_\beta(W)=K\,\underbrace{p^2(W-W_0)^2}_{\text{phase space}}\,F(Z,W)\,,
\end{equation}
where $W=E/(m_e c^2)+1$ and $W_0$ is the value of $W$ at the endpoint.
$K$ is a normalization constant. $F(Z,W)$ is the so called Fermi
function and given by
\begin{equation}
F(Z,W)=2 (\gamma+1)(2pR)^{2(\gamma-1)}e^{\pi\alpha Z W/p}\frac{\left|\Gamma(\gamma+i\alpha Z W /p)\right|^2}{\Gamma(2\gamma+1)^{2}}\quad\text{with}\quad \gamma=\sqrt{1-(\alpha Z)^2}\,,
\end{equation}
where $R$ the nuclear radius, which in itself has a dependence on $A$.
The Fermi function accounts for the fact that the outgoing electron is
moving in the Coulomb field of the nucleus and is derived from the
solution of the Dirac equation for a point-like and infinitely heavy
nucleus. The Fermi function is the leading order QED correction to
nuclear $\beta$-decay.

In equation~\ref{eq:simplebeta} the $\beta$-spectrum shape for allowed
decays is given, however fission fragments contain a significant
fraction of decay branches of unique and non-unique forbidden type.
For forbidden decays the simple phase space factor $p^2(W-W_0)^2$ is
multiplied by~\cite{Behrens:1969}
\begin{eqnarray}
p_\nu^2+p^2&\quad&1^\mathrm{st}\,\mathrm{unique}\,,2^\mathrm{nd}\,\text{non-unique}\nonumber\\
p_\nu^4+\frac{10}{3}p_\nu^2 p^2+p^4&\quad&2^\mathrm{nd}\,\mathrm{unique}\,,3^\mathrm{rd}\,\text{non-unique}\nonumber\\
p_\nu^6+7p_\nu^4 p^2+7p^4p_\nu^2+p^6&\quad&3^\mathrm{rd}\,\mathrm{unique}\,,4^\mathrm{th}\,\text{non-unique}\,.
\end{eqnarray}
We see, that the forbidden space phase factor is symmetric under the
exchange of $p$ and $p_\nu$ and thus, if we had $m_\nu=m_e$, the whole
expression would be symmetric between neutrino and electron energies,
too. If we keep the measured $\beta$-spectrum fixed and fit a
$\beta$-shape to it, any change to the $\beta$-shape which is
symmetric between $E_e$ and $E_\nu$ would not change the neutrino
spectrum. This the reason the corrections from forbidden decays to
inverted neutrino spectra have been found to be
small~\cite{Schreckenbach:1985ep,VonFeilitzsch:1982jw,Hahn:1989zr,Mueller:2011nm}
despite the overall large contribution of forbidden decays in fission
fragments.

It turns out, that for precision studies a number of additional
effects have to be taken into account and the modified
$\beta$-spectrum becomes~\cite{Wilkinson:1990}
\begin{equation}
\label{eq:completebeta}
N_\beta(W)=K\,p^2(W-W_0)^2\,F(Z,W)\,L_0(Z,W)\,C(Z,W)\,S(Z,W)\,G_\beta(Z,W)\,
(1+\delta_\mathrm{WM} W)\,.
\end{equation}
The neutrino spectrum is obtained by the replacements $W\rightarrow
W_0-W$ and $G_\beta\rightarrow G_\nu$. In the following we will
describe the physical origin and the actual expression used for each
of these additional terms. The size of the contribution of each of
these terms for a typical $\beta$-decay is shown in
figure~\ref{fig:betaspectrum}.

\subsection*{Finite size corrections}

There are two effects from the finite size of the nucleus: the
electric charge distribution is no longer point-like and the
hypercharge distribution is no longer point-like, {\it i.e.} the
nucleon is moving inside the nuclear potential. For all finite size
corrections we need to be able to determine the nuclear
radius\footnote{Differences due to different shapes of the charge
  distribution, {\it e.g.} constant within the nucleus or all charge
  on the surface are very small and can be neglected, see {\it
    e.g.}~\cite{Behrens:1982}.  Also, reasonable variations of the
  Elton formula have only a minor impact on our result, {\it i.e.} a
  variation of 10\% in $A$ will change the total flux by $0.05\%$.}
as a function of $A$ and we use the so called Elton
formula~\cite{Elton:1958}
\begin{equation}
R=0.0029 A^{1/3}+0.0063 A^{-1/3}-0.017 A^{-1}
\end{equation}
in units of $m_e c^2$. 

The electromagnetic finite size effect is
expressed by $L_0$ and we use the approximation given in
Ref.~\cite{Wilkinson:1990}
\begin{equation}
L_0(Z,W)=1+13\frac{(\alpha Z)^2}{60}-WR\alpha Z \frac{41-26\gamma}{15(2\gamma-1)}-\alpha Z R \gamma \frac{17-2\gamma}{30W(2\gamma-1)}+a_{-1}\frac{R}{W}+\sum_{n=0}^{5}a_n (WR)^n+0.41(R-0.0164)(\alpha Z)^{4.5}\,,
\end{equation}
where the $a_n$ are given by
\begin{equation}
a_n=\sum^{6}_{x=1}b_x (\alpha Z)^x\,,
\end{equation}
and the $b_x$ are taken from table 1 of Ref.~\cite{Wilkinson:1990} and
reproduced in table~\ref{tab:l0} in appendix~\ref{sec:sup}. We checked
that the error of this approximation in comparison with numerically
exact results given in~\cite{Behrens:1969} is smaller than $10^{-5}$.
Neglecting the last three terms in the above expression for $L_0$
still produces accurate results and this form of $L_0$ produces
results very close to the ones summarized as finite size effects in
equation~8 of Ref.~\cite{Mueller:2011nm}. 

The weak interaction finite size correction is of the following form
for Gamow-Teller decays\footnote{Almost all decays relevant in our
  problem are of this type and the corresponding expression for Fermi
  decays is numerically very close.}~\cite{Wilkinson:1990}
\begin{eqnarray}
C(Z,W)&=&1+C_0+C_1 W+C_2 W^2\quad\mathrm{with}\quad\\
C_0&=&-\frac{233}{630}(\alpha Z)^2-\frac{(W_0 R)^2}{5}+\frac{2}{35}W_0R\alpha Z\nonumber\,,\\
C_1&=&-\frac{21}{35}R\alpha Z + \frac{4}{9}W_0 R^2\nonumber\,,\\
C_2&=&-\frac{4}{9}R^2\,.\nonumber
\end{eqnarray}

In Ref.~\cite{Wilkinson:1974} it is estimated that the errors on both
these corrections, $L_0$ and $C$, are  smaller than 10\% of their size.
Thus, we can neglect this as a relevant source of error.

\subsection*{Screening correction}

\begin{figure}
\includegraphics[width=0.8\textwidth]{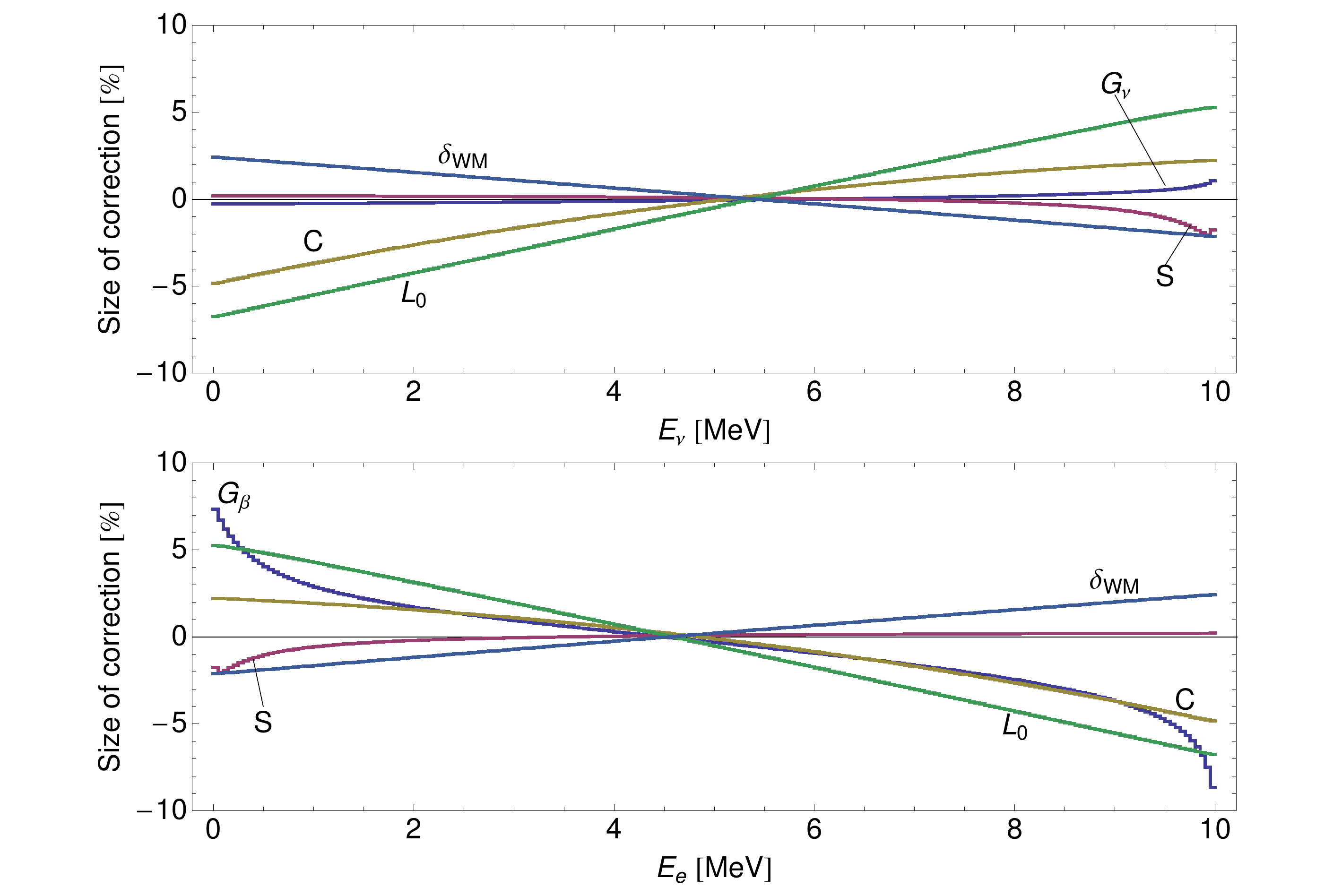}
\caption{\label{fig:betaspectrum}  (Color online) Shown is the relative size of the
  various corrections listed in equation~\ref{eq:completebeta} for a
  hypothetical $\beta$-decay with $Z=46$, $A=117$ and
  $E_0=10\,\mathrm{MeV}$. The upper panel shows the effect on the
  neutrino spectrum, whereas the lower panels shows the effect on the
  $\beta$-spectrum.}
\end{figure}

$S(Z,W)$ accounts for screening of the nuclear charge by all the
electrons in the atomic bound state and it effectively reduces the
charge ``seen'' be the outgoing electron. Our description of the
screening correction $S(Z,W)$ follows Ref.~\cite{Behrens:1982} and
we introduce the following quantities
\begin{equation}
\bar W=W-V_0\,,\quad \bar p = \sqrt{\bar W^2-1}\,,\quad  y=\frac{\alpha Z  W}{ p}\,\quad \bar y=\frac{\alpha Z \bar W}{\bar p}\,\quad \tilde Z = Z-1\,.
\end{equation}
$V_0$ is the so called screening potential and derived from numerical calculations and can be parametrized as
\begin{equation}
V_0=\alpha^2 \tilde Z^{4/3} N(\tilde Z)\,,
\end{equation}
and $N(\tilde Z)$ is taken as linear interpolation of the values in
table~4.7 of Ref.~\cite{Behrens:1982}, which is reproduced in table~\ref{tab:screening} in appendix~\ref{sec:sup}. Then, we obtain
\begin{equation}
S(Z,W)=\frac{\bar W}{W}\left(\frac{\bar p}{p}\right)^{(2\gamma-1)}
e^{\pi (\bar y-y)}\frac{\left|\Gamma(\gamma+i\bar y)\right|^2}{\Gamma(2\gamma+1)^{2}}\quad\text{for}\quad W>V_0\,,
\end{equation}
and $S(Z,W)=1$ for $W<V_0$. As can be seen from
figure~\ref{fig:betaspectrum} the effect of the screening correction
itself is rather small and the theory behind it is well
understood~\cite{Behrens:1982}, therefore we will not regard it a
source of theoretical errors.

\subsection*{Radiative corrections}

Radiative corrections are due to the emission of virtual and real
photons by the charged particles present in $\beta$-decay and it is
crucial to correctly account for both the virtual and real photon
contribution to ensure the proper cancellation of divergences. The
finite part of the virtual photon exchange contribution does depend on
the details of nuclear structure, but only affects the decay width and
not the spectral shape.  Therefore, the shape effect can be accurately
computed. In our presentation we will suppress all terms which only
affect the decay width, {\it i.e.} are independent of energy, and keep
only the energy dependent parts. The radiative correction at order
$\alpha$ has been computed by Sirlin~\cite{PhysRev.164.1767} for the
$\beta$-spectrum, with $\beta=p/W$
\begin{eqnarray}
g_\beta=3 \ln M_N -\frac{3}{4}+4\left(\frac{\tanh^{-1}\beta}{\beta}\right)\left(\frac{W_0-W}{3W}-\frac{3}{2}+\ln \left[2(W_0-W)\right]\right)+\frac{4}{\beta}L\left(\frac{2\beta}{1+\beta}\right)\nonumber\\
+\frac{1}{\beta}\tanh^{-1}\beta\left(2(1+\beta^2)+\frac{(W_0-W)^2}{6 W^2}-4\tanh^{-1}\beta\right)\,,
\end{eqnarray}
where $L(x)$ is the Spence function, defined as $L(x)=\int_0^x dt/t\,
\ln(1-t)$. The complete correction is then given by
\begin{equation}
G_\beta(Z,W)=1+\frac{\alpha}{2\pi}g_\beta\,.
\end{equation}

The radiative correction for neutrinos, $G_\nu$, has been derived in
Ref.~\cite{PhysRevD.52.5362} and we apply the form given in
Ref.~\cite{PhysRevD.52.5362} to the neutrino spectrum.  The
expressions are somewhat lengthy and involve integrals which can not
be solved in closed form, therefore we do not quote them here.  More
recently, the radiative correction for neutrinos has been derived in
closed form by Sirlin~\cite{Sirlin:2011wg}\footnote{I would like to
  thank T. Schwetz-Mangold for bringing this new work to my
  attention.} and we find that the energy dependence of the new
calculation, is numerically within 5\% of the previous
one~\cite{PhysRevD.52.5362}. Due to the fact that all spectra are
normalized to unity, differences in terms which do not depend on
energy will cancel. Thus. for our purposes the two results are
equivalent and since the new result~\cite{Sirlin:2011wg} is rather
compact, we quote it here
\begin{eqnarray}
h_\nu = 3 \ln M_N +\frac{23}{4}-\frac{8}{\hat\beta}L\left(\frac{2\hat\beta}{1+\hat\beta}\right)+8\left(\frac{\tanh^{-1} \hat\beta}{\hat\beta}-1\right)\ln (2\hat W \hat\beta)\nonumber\\+4\frac{\tanh^{-1}\hat\beta}{\beta}\left(
\frac{7+3\hat\beta^2}{8}-2\tanh^{-1}\beta\right)\,,
\end{eqnarray}
where $\hat\beta=\hat p/\hat E$ and $\hat p=\sqrt{\hat W^2-1}$ and
$\hat W=W_0-W_\nu$, where $W_\nu$ is the energy of the neutrino. The
neutrino radiative correction then becomes,
$G_\nu(Z,W_\nu)=1+\alpha/(2\pi)h_\nu$. Note, that the radiative
correction in the $\beta$-spectrum is quite large, whereas the
correction for the neutrino spectrum is very much smaller.

\subsection*{Weak magnetism}
\label{sec:wm}

The term ``weak magnetism'' was coined by Gell-Mann and he proposed it
as a sensitive test of the, then emerging, V-A theory of weak
interactions~\cite{GellMann:1958zz}. The weak magnetism contribution
to $\beta$-decay belongs to the class of so called induced currents,
{\it i.e.} currents which do not correspond to couplings present in
the initial Hamiltonian and they appear only at finite momentum
transfer.  Out of these induced currents, the weak magnetism term
typically yields the largest contribution to the shape of the
$\beta$-spectrum, see {\it e.g.}~\cite{Holstein:1974zf}, and therefore
we will only discuss this term. The effect of weak magnetism is to
modify the spectral shape of $\beta$-decay by a factor
\begin{equation}
  1+\delta_\mathrm{WM}W\quad\text{with}\quad \delta_{\mathrm{WM}}=\frac{4}{3}\underbrace{\frac{b}{M\,c}}_{=w}\,m_e\,,
\end{equation}
where $c$ is the Gamow-Teller matrix element, $b=\sqrt{2} \mu$, with
$\mu$ the magnetic transition moment and $M=A\,M_N$ the mass of the
nucleus.  Experimentally, the transition magnetic moment or $b$ can be
determined by measuring the magnetic dipole M1 $\gamma$ decay width,
$\Gamma_{\text{M}1}$, of the corresponding isovector transition of the
isobaric analog state and $b$ is then given by
\begin{equation}
\label{eq:CVC}
  b=\left(\frac{6\Gamma_{\text{M}1}\,M^2}{\alpha \,E_\gamma^3}\right)^\frac{1}{2}
\end{equation}
This derivation relies on the concept of conserved vector currents
(CVC), which is essentially a necessary result of gauge invariance.
In the so called impulse approximation, one further assumes that the
transition magnetic moment is entirely given by the intrinsic
anomalous magnetic moments of the proton, $\mu_p$, and neutron,
$\mu_n$, and one obtains for $w$, the following simple expression
\begin{equation}
w\simeq\frac{\mu_p-\mu_n}{M_N}\left|\frac{C_V}{C_A}\right|\,.
\end{equation}
Note, that in general there also would be a contribution to $b$ (or
$\mu$) from the orbital angular momentum of the decaying nucleon,
which we neglect here. Within the impulse approximation and neglecting
orbital angular momentum\footnote{Under these assumptions, there is
  also no contribution of the weak electric form factor, called $d$ in
  the notation of~\cite{Holstein:1974zf}, which could be sizable.},
there is no dependence on nuclear structure and therefore, $w$ has the
same value for all Gamow-Teller $\beta$-decays.  This approximation is
the basis for the statement in Ref.~\cite{Vogel:1983hi} that
$\frac{dN}{dE}=\frac{4}{3}w=0.5\%\,\mathrm{MeV}^{-1}$ is the universal
value for the weak magnetism slope parameter in all $\beta$ decays.
The validity of the necessary approximation seems to be supported by
the measured value for $\frac{dN}{dE}$ in the $A=12$ iso-triplet of
$^{12}$B, $^{12}$C and $^{12}$N~\cite{Wu:1977zz}.

At the same time, one can use the CVC hypothesis and use
equation~\ref{eq:CVC} to infer the weak magnetism slope parameter for
a number of isotopes from their measured $\gamma$-decay parameters. In
Ref.~\cite{Calaprice:1976} this analysis has been performed on a set
of 12 nuclei and a large variation of both $\frac{dN}{dE}$ and $b$ was
found, see table~11 of Ref.~\cite{Calaprice:1976}. In
table~\ref{tab:wm} we present a somewhat enlarged set of nuclei for
which we could identify the necessary information from the Evaluated
Nuclear Structure Data File (ENSDF) database~\cite{ensdf}. The
analysis follows the one outlined in Ref.~\cite{Calaprice:1976}. For
some of the nuclei we find slightly different results in comparison to
Ref.~\cite{Calaprice:1976}, which can be traced to changes in the
nuclear database and we have, in all cases, provided the corresponding
reference.

\begin{table}
\begin{tabular}{|rcr crr rrr rrc|}
\hline
\multicolumn{3}{|c}{decay} &$J_i\rightarrow J_f$&$E_\gamma$&$\Gamma_{M1}$&$b_\gamma$& ft& c &$b_\gamma/Ac$&$|dN/dE|$&Ref.\\
&&&&[keV]&[eV]&&[s]&&&$[\%\,\mathrm{MeV}^{-1}]$&\\
\hline

$^{6} \text{He}$&$\rightarrow$&$^{6} \text{Li}$&$0^+$$\rightarrow$$1^+$&$3563$&$8.2$&$71.8$&$805.2$&$2.76$&$4.33$&$0.646$&\cite{ensdf:1}\\
$^{12} \text{B}$&$\rightarrow$&$^{12} \text{C}$&$1^+$$\rightarrow$$0^+$&$15110$&$43.6$&$37.9$&$11640.$&$0.726$&$4.35$&$0.62$&\cite{ensdf:3}\\
$^{12} \text{N}$&$\rightarrow$&$^{12} \text{C}$&$1^+$$\rightarrow$$0^+$&$15110$&$43.6$&$37.9$&$13120.$&$0.684$&$4.62$&$0.6$&\cite{ensdf:2}\\
$^{18} \text{Ne}$&$\rightarrow$&$^{18} \text{F}$&$0^+$$\rightarrow$$1^+$&$1042$&$0.258$&$242.$&$1233.$&$2.23$&$6.02$&$0.8$&\cite{ensdf:4}\\
$^{20} \text{F}$&$\rightarrow$&$^{20} \text{Ne}$&$2^+$$\rightarrow$$2^+$&$8640$&$4.26$&$45.7$&$93260.$&$0.257$&$8.9$&$1.23$&\cite{ensdf:5}\\
$^{22} \text{Mg}$&$\rightarrow$&$^{22} \text{Na}$&$0^+$$\rightarrow$$1^+$&$74$&$0.0000233$&$148.$&$4365.$&$1.19$&$5.67$&$0.757$&\cite{mg22}\\
$^{24} \text{Al}$&$\rightarrow$&$^{24} \text{Mg}$&$4^+$$\rightarrow$$4^+$&$1077$&$0.046$&$129.$&$8511.$&$0.85$&$6.35$&$0.85$&\cite{al24}\\
$^{26} \text{Si}$&$\rightarrow$&$^{26} \text{Al}$&$0^+$$\rightarrow$$1^+$&$829$&$0.018$&$130.$&$3548.$&$1.32$&$3.79$&$0.503$&\cite{ensdf:6}\\
$^{28} \text{Al}$&$\rightarrow$&$^{28} \text{Si}$&$3^+$$\rightarrow$$2^+$&$7537$&$0.3$&$20.8$&$73280.$&$0.29$&$2.57$&$0.362$&\cite{al28}\\
$^{28} \text{P}$&$\rightarrow$&$^{28} \text{Si}$&$3^+$$\rightarrow$$2^+$&$7537$&$0.3$&$20.8$&$70790.$&$0.295$&$2.53$&$0.331$&\cite{al28}\\
\hline
$^{14} \text{C}$&$\rightarrow$&$^{14} \text{N}$&$0^+$$\rightarrow$$1^+$&$2313$&$0.0067$&$9.16$&$1.096\times 10^9$&$0.00237$&$276.$&$37.6$&\cite{ensdf:3}\\
$^{14} \text{O}$&$\rightarrow$&$^{14} \text{N}$&$0^+$$\rightarrow$$1^+$&$2313$&$0.0067$&$9.16$&$1.901\times 10^7$&$0.018$&$36.4$&$4.92$&\cite{Calaprice:1976}\\
$^{32} \text{P}$&$\rightarrow$&$^{32} \text{S}$&$1^+$$\rightarrow$$0^+$&$7002$&$0.3$&$26.6$&$7.943\times 10^7$&$0.00879$&$94.4$&$12.9$&\cite{ensdf:7}\\
\hline

\end{tabular}
\caption{\label{tab:wm} Gamow-Teller decays and the associated parameters needed for a computation of the weak magnetism slope parameter using the CVC hypothesis.}
\end{table}

This sample of nuclei is small, 13 nuclei only, and the majority is
much lighter than typical fission fragments, which raises the issue of
how representative this sample is for fission fragments. On the other
hand, these nuclei summarize our current experimental understanding of
weak magnetism. Setting these doubts aside, we can compute the average
value of $\frac{dN}{dE}$ and its standard deviation as a measure of
the error. In table~\ref{tab:wm} we have separated three isotopes based
on their unusually large $\log ft$ values. In those cases, the
Gamow-Teller matrix element is very small and thus the relative size
of the weak magnetism contribution is enhanced. As a result, all of
these 3 isotopes have large values of $\frac{dN}{dE}$.  Obviously,
these 3 large-$\log ft$ isotopes will completely dominate the average
and standard deviation and therefore, here the question of how
representative our sample is for fission fragments becomes acute. For
the main analysis presented in this paper, we therefore will exclude
these 3 isotopes and this yields the following mean and standard
deviations
\begin{equation}
\label{eq:dsde}
\frac{dN}{dE}=(0.67\pm 0.26) \,\%\,\mathrm{MeV}^{-1}\,,
\end{equation}
which is very close to the value obtained from using the
impulse approximation, but we also see that there is significant
variance. We will use the standard value of Ref.~\cite{Vogel:1983hi}
and assign a 100\% error to it.

If we include the large-$\log ft$ isotopes, the corresponding result
would be
\begin{equation}
\label{eq:dsdeH}
\frac{dN}{dE}==(4.78\pm 10.5) \,\%\,\mathrm{MeV}^{-1}\,,
\end{equation}
that is, a ten times larger mean with a many times larger relative error.
A shift of $\frac{dN}{dE}$ by $+0.5\,\%\,\mathrm{MeV}^{-1}$ causes a
shift of the neutrino rate of about $-1\%$. Thus, if one were to use
equation~\ref{eq:dsdeH} the resulting rate uncertainty would be about
20\% and the neutrino flux shift found in Ref.~\cite{Mueller:2011nm}
would be quite easily absorbed in this error bar. Even just a shift of
the central value is sufficient to bring the neutrino fluxes nearly
back to the original level. Obviously, large values of $\log ft$ imply
relatively long half-lives and long-lived isotopes play only a small
role in neutrino emission\footnote{{\it i.e.} above inverse
  $\beta$-decay threshold} from fission fragments. However, if the
endpoint energy is large, the relevant Gamow-Teller matrix element can
be quite small without leading to a very long lifetime. Also, it is
not clear how large weak magnetism would be in forbidden decays, since
there the leading order nuclear matrix element is already quite small
and thus all corrections tend to have relatively more weight. Thus,
the size of weak magnetism, or more generally induced currents, is the
one major source of theory uncertainty. In summary, the reactor
anomaly~\cite{Mention:2011rk} can be either attributed to new physics
in the form of a sterile neutrino or to a significant extent to some
not well understood nuclear physics. In particular, a detailed study
of the breakdown of the impulse approximation in large-$\log ft$
nuclei in the relevant mass range would be quite helpful, but is
beyond the scope of this paper.

\section{Extraction of the associated neutrino spectrum}
\label{sec:extract}

For a single $\beta$-decay branch the inversion from the
$\beta$-spectrum into the corresponding neutrino spectrum is
straightforward due to energy conservation, if one neglects recoil
effects, which are $\mathcal{O}(E_0/(A\,M_N))\sim10^{-4}$. In
particular, if the $\beta$-spectrum has been measured with good
precision, the neutrino spectrum can be known to the same level of
precision without any reliance on theory.  In practice, however, most
of the nuclei involved have many $\beta$-branches, most of which are
identified by $\gamma$-ray spectroscopy and not by direct measurement
of the $\beta$-spectrum; in these cases, $\beta$-decay theory as
outlined in section~\ref{sec:beta} plays a crucial role in obtaining
the associated neutrino spectrum. Of course, in a nuclear reactor a
very large number of different nuclei and $\beta$-branches contributes
to the neutrino spectrum, which makes the task of a first principles
calculation of the neutrino spectrum challenging and as we have
describe in section~\ref{sec:beta}, theory errors can play an
important role, even if one were to assume that the actual nuclear
data itself were flawless. The problems of this direct or {\it a
  priori} approach have been explored in detail in
Ref.~\cite{Mueller:2011nm}, with the conclusion that the inevitable
incompleteness of the nuclear databases makes it impossible to account
for all neutrinos and reliance on measured total $\beta$-spectra can
not be reduced beyond a certain level and eventually virtual
$\beta$-branches have to be used in addition to the {\it a priori}
computed spectra.

Neutrino spectra from reactor neutrinos have been obtained by theory
calculation by many
authors~\cite{PTP.45.1466,Davis:1979gg,Avignone:1980qg,Vogel:1980bk,Klapdor:1982sf,Klapdor:1982zz}
and for $^{238}$U this is still the only
possibility\footnote{According to K. Schreckenbach a measurement of
  the $^{238}$U $\beta$-spectrum is currently undertaken at the FRM II
  in Garching.}.  The neutrino spectra, which were the state of the
art prior to Ref.~\cite{Mueller:2011nm}, were obtained by the
exclusive use of virtual $\beta$-branches, which parameters were
determined by a fit to measured total
$\beta$-spectra~\cite{Schreckenbach:1985ep,VonFeilitzsch:1982jw,Hahn:1989zr}.
This procedure is \emph{not} independent of input from nuclear
databases, either. Here, the information from nuclear databases enters
in the form of the empirical mean proton number of the fission
fragments as a function of $E_0$, which we call the effective nuclear
charge, $\bar Z$, and is computed according to
\begin{equation}
\label{eq:zeff}
\bar Z(E_0)=\frac{\int_{E_0-\Delta/2}^{E_0+\Delta/2} dE' \eta(E') \, Z(E')}{\int_0^\infty dE' \eta(E')}\,,
\end{equation}
with $\Delta$ being an appropriately chosen energy interval.

From a mathematical point of view, the problem at hand is an inversion
or unfolding problem: Given the measured total $\beta$-spectrum
$\mathcal{N}_\beta$, one tries to infer the underlying distribution of
$\beta$-branches $\eta$, which we take to be a continuous distribution over all possible endpoints\footnote{For the moment we neglect the issue of forbidden decays.}, $E_0$
\begin{equation}
\label{eq:totspec}
\mathcal{N}_\beta(E_e)=\int dE_0 N_\beta(E_e,E_0;\bar Z) \,\eta(E_0)\,.
\end{equation}
If we neglect the dependence on $\bar Z$, problems like this one are
known as Fredholm integral equation of the first kind. These problems
are ill-posed and in general have no unique solution, therefore it is
necessary to provide an additional constraint, a so called
regularization scheme. The choice of regularization scheme is
essentially an art and not a science, since one has to find the
appropriate compromise between finding a reasonably stable solution
and the introduction of an unacceptable bias into the result; for an 
introduction into the topic of regularized inversion see {\it e.g.}
Ref.~\cite{Craig:1986}. The regularization scheme, which has been
developed for our case, relies essentially on averaging the solution
over a finite range of $E_\nu$, which corresponds to a top-hat filter.
This procedure has been studied in great detail in~\cite{Vogel:2007}
based on a synthetic data set. The use of a synthetic data set as
proxy for real data allows to quantify the size of the bias and thus
serves as a valuable tool to fine tune the regularization scheme. In
Ref.~\cite{Mueller:2011nm} it was shown that a $\beta$-spectrum
computed from cumulative fission decay yields and the ENSDF database
reproduces the measured total $\beta$-spectrum for $^{235}$U within an
error of about 10\% and therefore can serve as an accurate proxy for
the actual data set. We will use a total $\beta$-spectrum derived from
the endpoints contained in the ENSDF database\footnote{The endpoints,
  branching fractions, the assignment of the degree of forbiddeness and
  simple $\beta$-spectra which were computed according to
  Ref.~\cite{Mueller:2011nm}, {\it i.e.} without the terms $C$, $S$,
  $G_\nu$, were kindly provided in machine readable format by the
  authors of Ref.~\cite{Mueller:2011nm}.} and we use fission yields
for fission by $25\,\mathrm{meV}$, {\it i.e.} thermal, neutrons from
the JEFF database, version 3.1.1~\cite{JEFF}. Note, that we neglect
the contribution from epithermal and fast neutrons which would change
the total spectrum by only a few percent and therefore, does not
affect its suitability or accuracy as proxy. We also do not apply the
small correction for finite irradiation time, this correction can be
found in table~VII of Ref.~\cite{Mueller:2011nm}. In computing the
resulting $\beta$-spectra we apply the results obtained in
section~\ref{sec:beta}.  We use this data set, which consists of
approximately 550 isotopes and contains about 8\,000 individual
$\beta$-branches, to perform a critical study of an inversion purely
based on virtual branches.

In Ref.~\cite{Vogel:2007} it was shown that given a precise enough
measurement of the total $\beta$-spectrum and accurate knowledge of
$\bar Z$ the true neutrino spectrum could be recovered to better than
within 1\%. Unfortunately, several of the assumptions made in this
analysis do not apply here: first, the available data was taken with
a resolution of $50\,\mathrm{keV}$ and secondly $\bar Z$ can be known
only approximately, since the underlying database is incomplete,
otherwise there would be no need for the use of virtual branches.
Also, the analysis in Ref.~\cite{Vogel:2007} does not include the
various corrections to a allowed $\beta$-decay shape and the effect of
statistical fluctuations of the actual $\beta$-data was not taken into
account. In the following we will quantify the impact of this various
factors.

\subsection{Bias \& Statistical Errors} 
\label{sec:bias}

\begin{figure}[t]
\includegraphics[width=\textwidth]{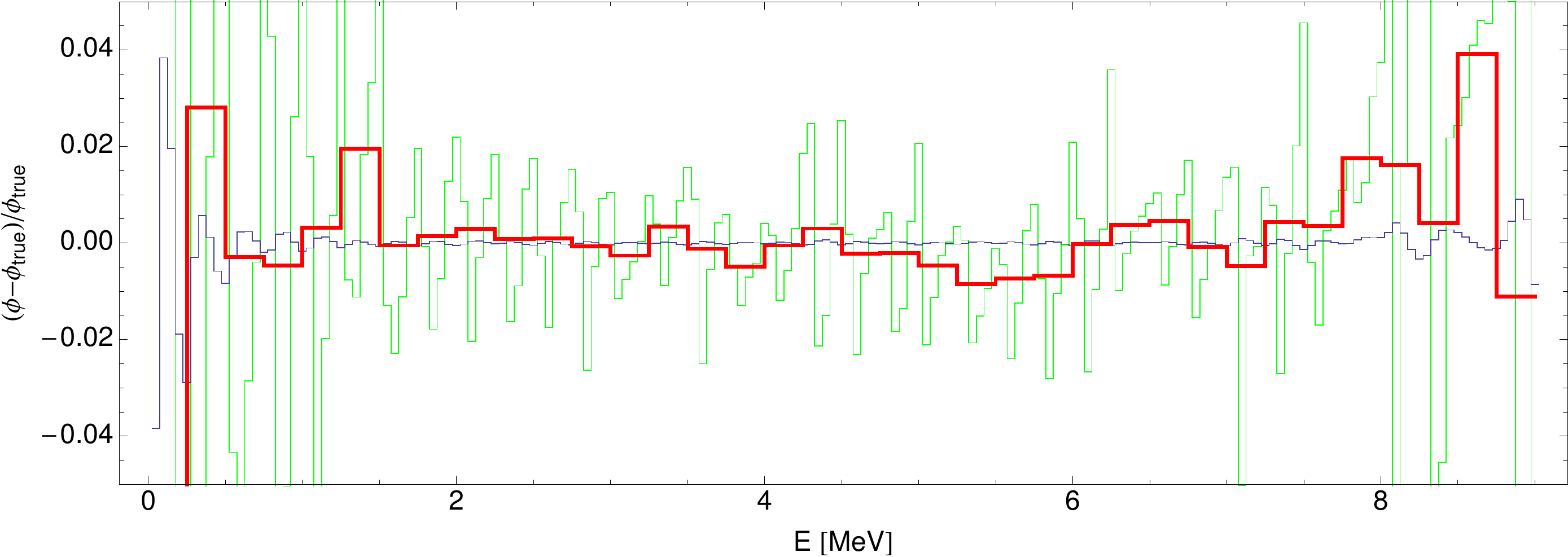}
\caption{\label{fig:inverse} (Color online) The basic inversion
  procedure, shown for a synthetic data set for $^{235}$U. The green
  (thin, gray) line shows the neutrino residuals in $50\,$keV bins and
  the blue (thin, black) line shows the $\beta$-residuals in
  $50\,$keV bins. The thick red line depicts the neutrino residuals in
  $250\,$keV bins.}
\end{figure}

The basic procedure is to take the measured $\beta$-spectrum, for
which, at this stage, we will substitute our simulated
$\beta$-spectrum and to start from the highest energy data point
$n_\beta^N$, where $N$ is the number of data points. Then, one fixes a
certain size $s$ of a slice $S$ and takes the data points
$S_1=\{n_\beta^{N-s},n_\beta^{N-s+1},\ldots,n_\beta^{N}\}$. We then
fit an allowed $\beta$-spectrum with a free endpoint $E_1$ and
amplitude $a_1$ to the data contained in $S_1$. We continue this
$\beta$-branch to all energies and subtract it from the remaining
$N-s$ data points.  This procedure is repeated till all data points
have been fitted and as result we have a set of $v=N/s$ endpoints and
amplitudes $\{E_i,a_i\}$, where $v$ is the number of virtual branches.
The set $\{E_i,a_i\}$ is a discrete approximation to $\eta(E_0)$ and
can be used to compute the neutrino spectrum, by trivially inverting
each virtual $\beta$-branch into its corresponding neutrino spectrum.
The result of this procedure for $^{235}$U with ``data'' in bins of
$50\,$keV and $s=5$, corresponding to 30 virtual branches, is shown as
green (thin, gray) line in figure~\ref{fig:inverse}. The thin blue (black)
line shows the $\beta$-residuals, which are obviously very small.

The $50\,$keV anti-neutrino residuals show oscillations which are due
to the simple fact that the $\beta$-spectrum is non-zero at the
endpoint. The amplitude of these oscillations in increased by the
inversion procedure and this increase is a direct consequence of the
problem being ill-posed in the mathematical sense. To obtain a more
regular solution, we can average our result and obtain the thick red
line by averaging over $250\,$keV bins. Clearly, this regularized
solution shows oscillations of much smaller amplitude, especially in
the region of interest between 2 and 8\,MeV. We also observe that the
remaining oscillations do not average to zero, {\it e.g.} between 4
and 6\,MeV both, the regularized and non-regularized, solutions
predict a neutrino flux which is too low by about 1\%; the result is
biased. A finite bias is not surprising and quite common in this type
of inversion problem. Using a synthetic data set it is straightforward
to quantify the bias by comparing the result of the inversion with the
known true flux, as done in figure~\ref{fig:inverse}. We, also, can
vary the starting point of the inversion by removing an increasing
number of consecutive data data points at the high energy end and
repeat the procedure. The overall features stay the same, but the
phase of the oscillations of the non-regularized solution changes
gradually and once we have removed $s$ data points we obtain the
initial solution again. This variation of the phase introduces a
slight spread in the magnitude of the bias, which we will assign as
error of the bias correction.

\begin{figure}[t]
\includegraphics[width=\textwidth]{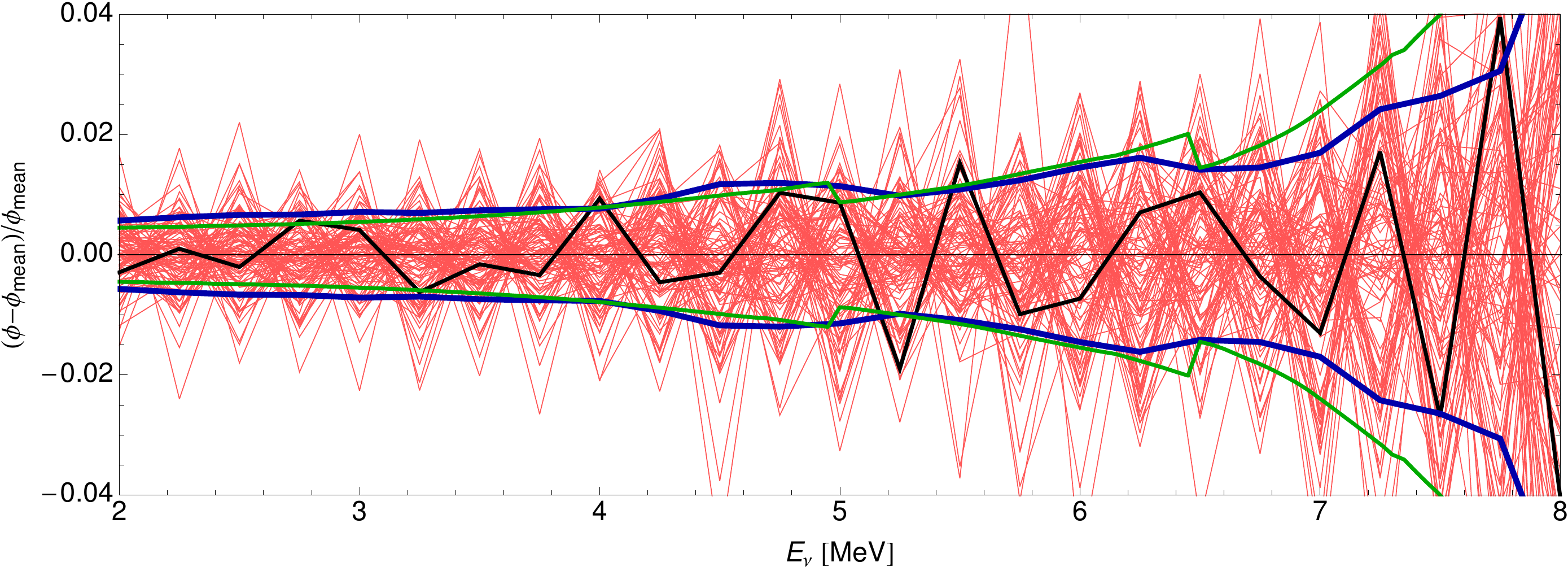}
\caption{\label{fig:correlation} (Color online) The neutrino flux from
  the inversion of 1000 random realizations of a synthetic
  $\beta$-spectrum for $^{235}$U relative to the mean outcome of these
  1000 trials. The red (thin, gray) lines show a subset of 100 trials.
  The thick, ragged black line shows one particular example.  The
  smooth thick, blue (black) lines show the standard deviation in each
  bin, which is the same as the square root of the diagonal elements
  of the covariance matrix. The dark green (thick, dark gray) lines
  are the statistical error of the $\beta$-spectrum scaled up by a
  factor of 7.}
\end{figure}
The oscillations apparent in figure~\ref{fig:inverse} show that the
inversion procedure amplifies small perturbations and may be not
stable with respect to statistical fluctuations, which in real data
are inevitable. Given the statistical errors of the $\beta$-spectra we
can add random fluctuations with the same variance to our synthetic
$\beta$-spectra and for each realization of a randomized
$\beta$-spectrum we can perform an inversion. We have done this for
1000 random data sets and the results are shown as red (thin, gray)
lines in figure~\ref{fig:correlation}, where we plot the relative
deviation from the mean. The thick black lines shows one
characteristic example.  It is quite obvious that there are very
strong anti-correlations between neighboring bins. These
anti-correlations are due to oscillations in the inversion procedure,
which are excited by statistical fluctuations.  Since, these
oscillations are unphysical and merely a feature of the inversion
procedure, we do not want to propagate these oscillations into the
final result. The thick smooth, black line shows the standard
deviation in each bin, which turns out to be the same as the square
root of the corresponding diagonal element of the covariance matrix of
the 1000 randomized inversions.  The range delimited by the standard
deviation is a good estimate of the statistical error, therefore, we
will drop all off-diagonal elements of the covariance matrix. The
resulting errors are found to be several times larger in the inverted
neutrino spectrum than they are in the underlying $\beta$-spectrum.
The errors of the $\beta$-spectrum are shown as dark green (thick,
gray) lines but multiplied by a factor of 7. Note, that this scaling
factor is not universal and depends sensitively on the details of the
inversion, like bin width.

\subsection{Effective average nuclear charge}
\label{sec:zeff}

\begin{figure}[t]
\includegraphics[width=\textwidth]{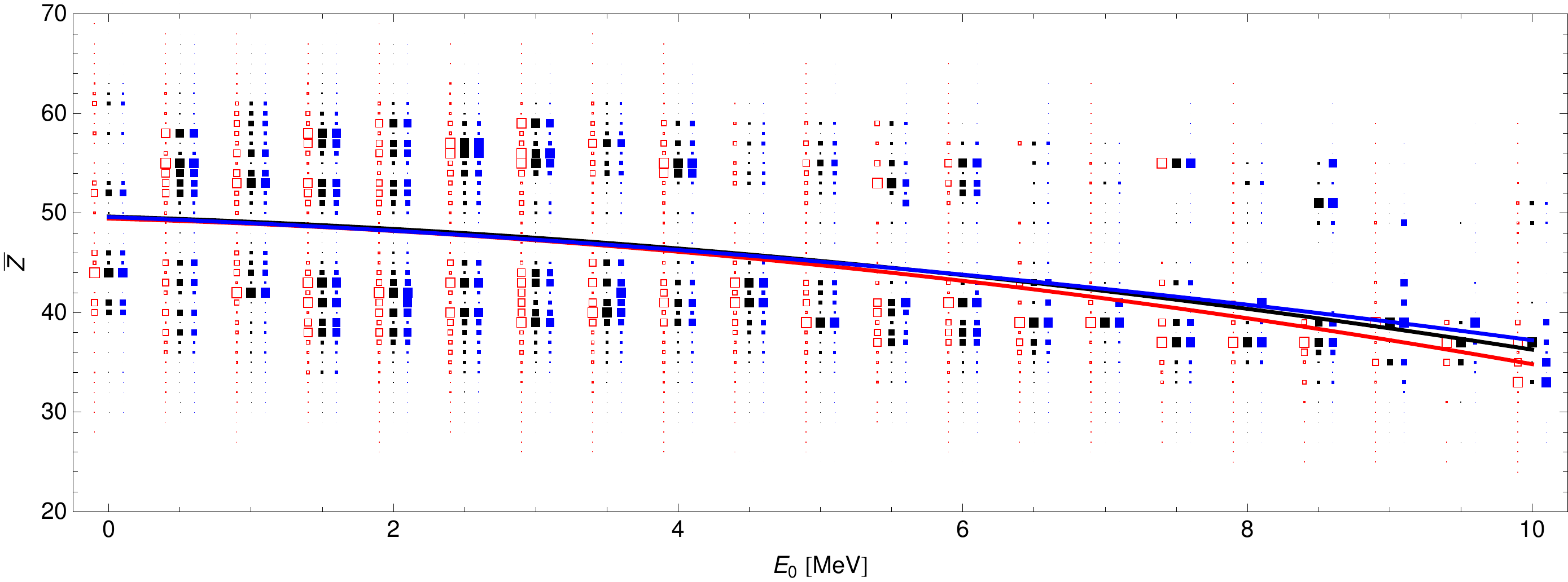}
\caption{\label{fig:Zeff} (Color online) The effective nuclear charge
  $\bar Z$ of the fission fragments of $^{235}$U as a function of
  $E_0$. The area of the each box is proportional to the contribution
  of that particular $Z$ to the fission yield in that energy bin. The
  lines are fits of quadratic polynomials: black -- ENSDF database,
  blue (dark gray) -- adding those isotopes missing from ENSDF by
  assuming that there is only one $\beta$-branch, each with
  $E_0=Q_\beta$, red (light gray) -- maximum pandemonium as defined in
  the text.  The blue (dark gray, rightmost), filled boxes show the
  resulting distribution of adding the missing isotopes with
  $E_0=Q_\beta$. The red (light gray, leftmost), empty boxes show the
  distribution in the maximum pandemonium approximation.}
\end{figure}

Finally, we study the degree to which $\bar Z$ can be determined and
how the incompleteness of nuclear databases affects the inversion
procedure. We fit a second order polynomial in $E_0$, where we weight
each $\beta$-decay branch by it fission yield $Y_{A,Z}$ and the
corresponding branching ratio $b^i_{A,Z}$, the result is shown as a
black line in figure~\ref{fig:Zeff}. The black boxes are a
two-dimensional histogram of the $Z$ distribution of the fission
fragments using the ENSDF database. Note, that this distribution is
bi-modal, because fission proceeds into two asymmetric fragments, and
it is therefore, quite unexpected that a simple energy-dependent mean
($\bar Z$) actually is sufficient to compute accurate neutrino
spectra, as previously shown in~\cite{Vogel:2007} and in
figure~\ref{fig:inverse}.

We know from comparing the fission yields with the ENSDF database how
many and which nuclei are missing and although we have no detailed
information on $\beta$-branches for those nuclei, ENSDF does contain
the $Q_\beta$ for all of the missing nuclei. Thus, we can try to
bracket the effect of the missing nuclei by computing $\bar Z$ under
two different approximations.  First, we assume there is indeed only
one $\beta$-branch with $E_0=Q_\beta$ for each missing isotope. In
this case, the effect of the missing isotopes is to add to the high
energy part and increase $\bar Z$ by a rather small amount, as shown
by the blue (dark gray) curve in figure~\ref{fig:Zeff}. In nuclei far
from stability, there will be very many, but individually quite weak
$\beta$-branches, which makes it very difficult to infer the correct
level scheme and $\beta$-decay branching fractions from
$\gamma$-spectroscopy, this is termed as pandemonium
effect~\cite{Hardy:1977}. In Ref.~\cite{Mueller:2011nm} the
pandemonium effect is taken care of by replacing about 50-100 of the
nuclei in the ENSDF database by data which was obtained specifically
avoiding the pitfalls of the pandemonium. Here, we would like to use
an approximation to this method, which we will call the ``maximum
pandemonium''. We take the $17.5\%$ most neutron rich
nuclei\footnote{$17.5\%$ was chosen because the resulting deficit in
  the total $\beta$-spectra is very similar in magnitude and shape to
  the one shown in figure~5 of Ref.~\cite{Mueller:2011nm}.}  from the
ENSDF database and all the $Q_\beta$ values of the missing nuclei and
replace all their $\beta$-spectra using the following algorithm: We
distribute the $\beta$-decay strength evenly between $0$ and $Q_\beta$
using about 10 decay branches. The result is shown as red (light gray)
line in figure~\ref{fig:Zeff}. This should yield a reasonable
approximation, at least when one averages over a large number of
nuclei, as we presently do. The maximum pandemonium approximation is
of course rather crude and would be not sufficient for a direct {\it a
  priori} calculation of the neutrino spectrum, but it allows us to
gauge the magnitude of the effect of the incompleteness of the nuclear
database we use. We repeat this exercise for the other isotopes as
well and the results on $\bar Z$ are summarized in
table~\ref{tab:Zeff}.
\begin{table}[h]
\begin{tabular}{|c|ccc|}
\hline
Isotope& $c_0$ &$ c_1$ & $c_2$\\
\hline
$^{235}$U&$48.992^{+0}_{-0.164}$&$-0.399^{+0.161}_{-0}$&$-0.084_{-0.044}^{+0}$\\
$^{239}$Pu&$49.650_{-0.214}^{+0}$&$-0.447^{+0.036}_{-0}$&$-0.089_{-0.016}^{+0}$\\
$^{241}$Pu&$49.906_{-0.178}^{+0}$&$-0.510^{+0.160}_{-0}$&$-0.044_{-0.052}^{+0}$\\
\hline
\end{tabular}
\caption{\label{tab:Zeff} The coefficients for the parametrization of $\bar Z$ for the various isotopes. Also given are the range as obtained from the maximum pandemonium approximation. $c_i$ is the coefficient of $E_0^i$ in a second order polynomial in $E_0$. Note, that in our definition $\bar Z$ is the charge of the parent nucleus.}
\end{table}
The induced systematic error is correlated between the isotopes since
it has a common physical cause.  Note, that this is the only place at
which information from nuclear databases enters the computation of the
neutrino spectrum using only virtual $\beta$-branches. In principle,
the nuclear mass $A$ will show a similar change as a function of
$E_0$, but variations of $A$ have a much smaller impact on the
neutrino spectrum and therefore fixing the effective nuclear mass,
$\bar A$, at $\bar A=117$ does not lead to any relevant additional
bias or error. The value for $\bar A$ is chosen based on the average
nuclear radius, where the average is determined by weighting each
isotope with its cumulative fission yield and we find a range of about
116-118 for $^{235}$U to $^{241}$Pu.

\section{Application to $^{235}$U, $^{239}$Pu and $^{241}$Pu} 
\label{sec:results}

In this section, we will apply the results obtained so far to a direct
inversion of the $\beta$-spectra which were measured in the 1980s at
ILL. We use the $\beta$-spectra for $^{235}$U as presented in
Ref.~\cite{Schreckenbach:1985ep}, for $^{239}$Pu from
Ref.~\cite{VonFeilitzsch:1982jw,Hahn:1989zr}\footnote{The actual data
  on electron spectra is the same in both sources, however the
  neutrino spectrum was re-derived in the later reference using
  updated nuclear databases.} and for $^{241}$Pu from
Ref.~\cite{Hahn:1989zr}.  The original data for all 3 isotopes was
recorded in 50\,keV bins, but unfortunately was published only in
250\,keV bins, which is not sufficient for our purposes. We, however,
were able to obtain finer binned data from one of the
authors~\cite{Schreckenbach:2011} of
Refs.~\cite{Schreckenbach:1985ep,VonFeilitzsch:1982jw,Hahn:1989zr}:
for $^{235}$U we have data in 50\,keV bins from 1.5 to 9.5\,MeV, for
$^{239}$Pu we have data in 100\,keV bins from 1 to 8\,MeV and for
$^{241}$Pu we have data in 100\,keV bins from 1.5 to 9\,MeV. Thus, we
have data for all 3 isotopes only in the energy interval between 1.5
and 8\,MeV and, since the threshold for inverse $\beta$-decay is
1.8\,MeV and the results in Ref.~\cite{Mueller:2011nm} range from 2 to
8\,MeV, we will present result only for this interval. Together with
the original data, also came the size of the statistical error in each
bin, which when summed up over 250\,keV bins agree with the values
published in
Refs.~\cite{Schreckenbach:1985ep,VonFeilitzsch:1982jw,Hahn:1989zr}, as
do the data itself. Note, that the data on $^{235}$U not only has the
smallest bin size but also the smallest statistical errors, whereas
$^{239}$Pu has the largest errors. There  is a calibration or
normalization error for each data set and we take these values
directly from
Refs.~\cite{Schreckenbach:1985ep,VonFeilitzsch:1982jw,Hahn:1989zr} and
list them in column~9 of tables~\ref{tab:u235}-\ref{tab:pu241}.

For each isotope we perform the inversion as outlined in
section~\ref{sec:extract} but now for the actual data, we use 30
virtual branches with a slice size of 5 for $^{235}$U and 23 virtual
branches for $^{239}$Pu and 25 for $^{241}$Pu both with a slice size
of 3. We have tested that variations of the slice size around the
values given only mildly affect the results and the chosen values
provide the smallest errors around 4\,MeV. Also, with these choices
the energy range of a slice is about the same in all 3 isotopes and is
close to 250\,keV. By using synthetic data sets for each isotope we
determine the bias and statistical errors, the later is shown column 5
of tables~\ref{tab:u235}-\ref{tab:pu241}, where the square root of the
diagonal elements of the covariance matrix are given.  We apply the
resulting bias correction, column 3 of
tables~\ref{tab:u235}-\ref{tab:pu241}, to the results extracted from
the actual data and assign the spread of biases obtained by varying
the starting point of the inversions as error, which is given in
column 6 of tables~\ref{tab:u235}-\ref{tab:pu241}, see also
section~\ref{sec:bias}. Then we perform an inversion using the upper
and lower ends of the ranges for $\bar Z$ from section~\ref{sec:zeff}
and for the weak magnetism correction from section~\ref{sec:wm}. The
relative differences to the initial inversion result are listed as
errors in columns~7 and~8 of tables~\ref{tab:u235}-\ref{tab:pu241}.
Finally the error components in columns~5-9 are added in quadrature
and quoted as total error in column~10 of
tables~\ref{tab:u235}-\ref{tab:pu241}.

The total error quoted in tables~\ref{tab:u235}-\ref{tab:pu241} is
only an indication of the actual errors since there are correlations
between the various bins. The statistical error is uncorrelated
between bins as well as the bias error. Both the errors due to $\bar
Z$ and weak magnetism (WM) are fully correlated between all bins and
all isotopes.  The normalization errors is fully correlated between
bins and between isotopes, since they same apparatus and method to
determine the normalization was used for all 3 isotopes. Due to the
asymmetric nature of the errors from $\bar Z$ and weak magnetism (WM)
they can not be included exactly in a total covariance matrix.
However, already the combined errors form $\bar Z$ and WM are close to
symmetric and the overall errors are very close to symmetric, thus it
is a reasonable approximation to neglect the small asymmetry in errors
for most applications.

For completeness, we also provide a parametrization of our results
using the familiar exponential of a 5$^\mathrm{th}$ order
polynomial~\cite{Huber:2004xh,Mueller:2011nm}:
\begin{equation}
\phi_\nu(E_\nu)=\exp\left(\sum_{i}^6 \alpha_i E_\nu^{i-1}\right)\,.
\end{equation}

In performing the fit we follow the description given in
Refs.~\cite{Huber:2004xh,Mueller:2011nm} but we do not include
contributions from the error on $\bar Z$ and WM since they are
correlated between isotopes and do not change the fit appreciably. The
resulting best fit parameters and the minimum $\chi^2$ values are
given in table~\ref{tab:param}.

\begin{table}[h!]
\begin{tabular}{|c|c|ccc ccc|}
\hline
Isotope&$\chi^2_\mathrm{min}$&$\alpha_1$&$\alpha_2$&$\alpha_3$&$\alpha_4$&$\alpha_5$&$\alpha_6$\\
\hline
$^{235}$U&49.3&$4.367$&$-4.577$&$2.100$&$-5.294\times 10^{\text{-1}}$&$6.186\times 10^{\text{-2}}$&$-2.777\times 10^{\text{-3}}$\\
\hline
$^{239}$Pu&20.8&$4.757$&$-5.392$&$2.563$&$-6.596\times 10^{\text{-1}}$&$7.820\times 10^{\text{-2}}$&$-3.536\times 10^{\text{-3}}$\\
\hline
$^{241}$Pu&15.&$2.990$&$-2.882$&$1.278$&$-3.343\times 10^{\text{-1}}$&$3.905\times 10^{\text{-2}}$&$-1.754\times 10^{\text{-3}}$\\
\hline
\end{tabular}
\caption{\label{tab:param} Result of a fit of a 5$^\mathrm{th}$ order polynomial to the logarithm of the flux. The number of degrees of freedom is $25-6$.}
\end{table}

Obviously, the fit for $^{235}$U is quite bad, with a $\chi^2/$dof of
more than 2. Also, for all 3 isotopes the fit parameters are highly
correlated and we therefore do not provide any fit errors or
correlation matrices, since it seems doubtful whether these could be
used to model the errors in the underlying neutrino fluxes. This
parametrization should not be used for actual data analysis or error
propagation\footnote{Nonetheless, this parametrization can be safely used
  to extrapolate our results to higher energies, since there,
  the errors of the actual fluxes are sufficiently large to render the
  inaccuracies of this parametrization harmless.}. Instead, we
recommend to rebin our results by using linear interpolation and
integrating the resulting fluxes over the new bins.  In the same
fashion the errors can be rebinned and we provide results for 250\,keV
and 50\,keV bins in machine readable format~\cite{Huber2011}.

\section{Discussion}
\label{sec:discussion}

\begin{figure}
\includegraphics[width=\textwidth]{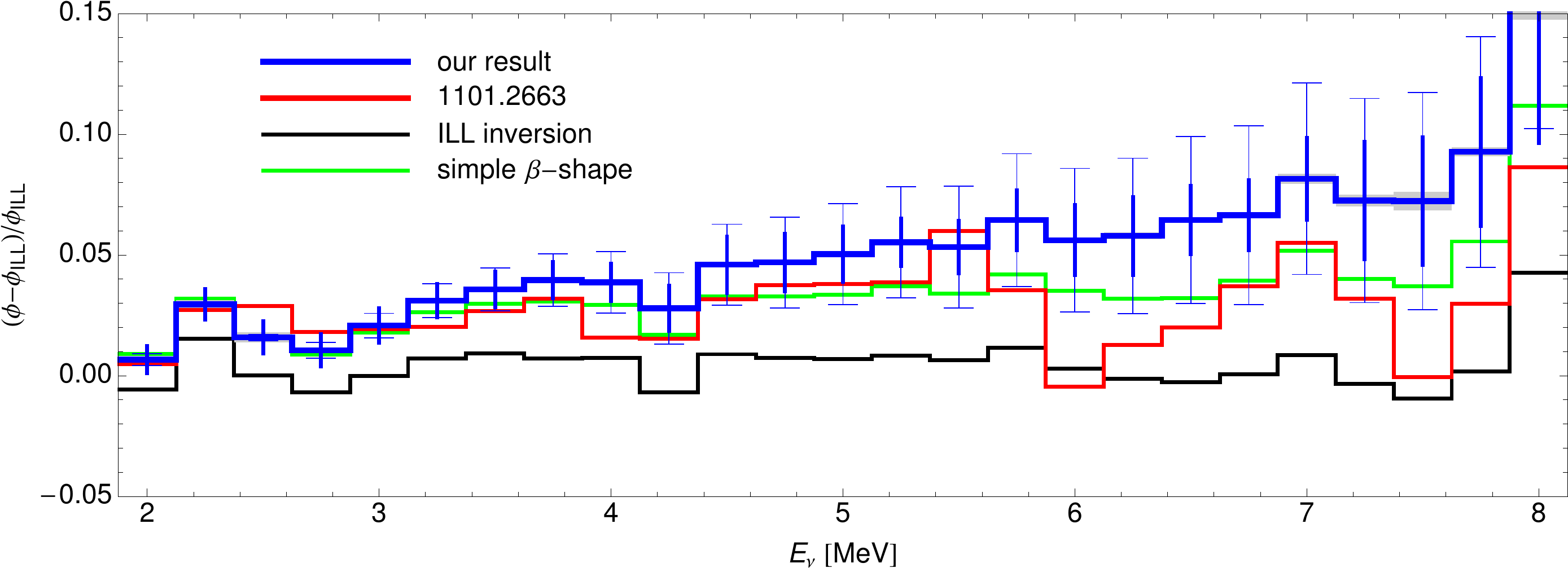}
\caption{\label{fig:u235final} (Color online) Comparison of our result
  for $^{235}$U with previous inversions, labeled ILL for the results
  from Ref.~\cite{Schreckenbach:1985ep} and labeled 1101.2663 for the
  results from Ref.~\cite{Mueller:2011nm}. The thin error bars show
  the theory errors from the effective nuclear charge $\bar Z$ and
  weak magnetism. The thick error bars are the statistical errors,
  whereas the light gray boxes show the error from the applied bias
  correction. The green line, referred to as simple, shows the result,
  if we use the same description of $\beta$-decay as in
  Ref.~\cite{Mueller:2011nm}. The black line, referred to as ILL
  inversion, shows our result if we completely follow the procedure
  outlined in Ref.~\cite{Schreckenbach:1985ep}, including their
  effective nuclear charge. }
\end{figure}
The $\beta$-spectra from
Refs.~\cite{Schreckenbach:1985ep,VonFeilitzsch:1982jw,Hahn:1989zr}
have been previously inverted into neutrino spectra and we, therefore,
will start by comparing our result with previous
results~\cite{Mueller:2011nm,Schreckenbach:1985ep,VonFeilitzsch:1982jw,Hahn:1989zr}.
In figure~\ref{fig:u235final} we show our result (thick blue/black
line) relative to the results presented in
Ref.~\cite{Schreckenbach:1985ep}, which is denoted as
$\phi_\mathrm{ILL}$. We clearly observe that our results point to a
significantly enhanced flux with respect to $\phi_\mathrm{ILL}$. The
results from Ref.~\cite{Mueller:2011nm} are shown as red (thin, dark
gray) line and up to about 5.5\,MeV there is good agreement, whereas
at higher energies the result of Ref.~\cite{Mueller:2011nm} exhibits
strong oscillation relative to both $\phi_\mathrm{ILL}$ and our
result, which may be due the coarser binning of the $\beta$-spectra
used in Ref.~\cite{Mueller:2011nm}. The thin error bars are the sum of
the theory errors, {\it i.e.} the effective nuclear charge $\bar Z$
and weak magnetism WM, the thick error bars show the statistical
errors and finally the gray boxes show the error from the bias
correction. The theory errors are correlated between all bins, the
statistical and bias correction error are uncorrelated. Not shown is
the, common to all data sets, normalization error. The thin green line
shows our result, if we use the same, simplified, description of the
$\beta$-decay spectrum as in Ref.~\cite{Mueller:2011nm} and we see
that it is responsible for a sizable fraction of the difference of our
result with the others at energies above 5\,MeV. Finally, the thin black
line shows our result, if we follow the procedure outlined in
Ref.~\cite{Schreckenbach:1985ep}, which in particular includes their
form of $\bar Z$ and the fact, that all corrections to a allowed
$\beta$-spectrum shape are applied in an average fashion and we
recover the original result quite well. Similar figures can be made
for the other isotopes and the results, generally, would be similar,
too; the error bars would be much larger due to the 100\,keV binning
of the original data and the larger statistical errors in the
$\beta$-spectra. As far as the origin of the upward shift in flux
relative to the original
results~\cite{Schreckenbach:1985ep,VonFeilitzsch:1982jw,Hahn:1989zr}
is concerned, we agree at some level with the comments made in
Ref.~\cite{Mueller:2011nm}, where at energies below 4\,MeV the main
effect is ascribed to the branch-by-branch implementation of the
various corrections to a allowed $\beta$-spectrum shape. We would like
to point out that both an implementation in each virtual branch as
well as an implementation in each physical branch in an {\it a priori}
calculation yield the same effect. Moreover, it is likely that a newly
derived average correction based on modern nuclear databases maybe
nearly as accurate as a branch-by-branch implementation. Of course, to
derive this new average correction would presumably imply to actually
study the corrections first branch-by-branch, so nothing would be
gained. At energies above 5\,MeV, the new form of the effective
nuclear charge $\bar Z$ as given in table~\ref{tab:Zeff} does make up
a large fraction of the shift. Note, that in contrast to
Ref.~\cite{Mueller:2011nm} we do find that an effective nuclear charge
is sufficiently accurate and the effect of the actual distribution of
$Z$ can be approximated quite well by its, properly defined, average;
this conclusion is supported by the results of Ref.~\cite{Vogel:2007}.
At high energies the contribution from the wave function convolution
and screening correction on the $\beta$-shape contribute significantly
to the overall shift. Once we combine all factors we find about the
same upward shift in all isotopes, however we would not characterize
it as constant over all energies.

In table~\ref{tab:comparison} we quantify the mutual agreement or
disagreement of the various fluxes and associated inverse
$\beta$-decay event rates, where we use the cross section
from~\cite{Vogel:1999zy}. Inverse $\beta$-decay is frequently used for
the detection of reactor anti-neutrinos. In this table we list for
each isotope the total difference over all energy bins from $2-8$\,MeV
of the fluxes from
Refs.~\cite{Schreckenbach:1985ep,VonFeilitzsch:1982jw,Hahn:1989zr}
labeled $\phi_\mathrm{ILL}$ and the ones from
Ref.~\cite{Mueller:2011nm} labeled $\phi_\mathrm{MLF}$ to our results.
We provide the total error as well as the individual contributions
from statistics, theory and the bias correction. The ratio of the
total difference to the total error is what we call the ``rate
significance'', {\it i.e.} the significance of the total flux change.
We find the total flux change relative to $\phi_\mathrm{ILL}$ to be in
the range $+(2.4-3.2)\%$. The upward shift in fluxes is significant at
the $(3.7-4.0)\,\sigma$ level.  This shift is slightly larger than the
one found in Ref.~\cite{Mueller:2011nm}, however if we compare our
total flux with respect to $\phi_\mathrm{MLF}$ we find very small
differences for $^{235}$U and $^{241}$Pu.  For $^{239}$Pu the
situation is different and we have a lower total flux than
$\phi_\mathrm{MLF}$, which however is entirely due to the first energy
bin and could be due a spurious oscillation in the inversion procedure
in either analysis.  If we perform the same comparisons for inverse
$\beta$-decay event rates, rows labeled as $R_\mathrm{ILL}$ and
$R_\mathrm{MLF}$, we find that in total event rates the upward shift
with respect to $R_\mathrm{ILL}$ is a bit larger than for the fluxes
with $+(3.7-4.7)\%$, which is due to the relatively large shift at
high energies which is enhanced by the $E_\nu^2$ energy dependence of
the cross section. These rates shifts are significant at
$(2.4-3.0)\,\sigma$ and the reason for the significance being smaller
than for the fluxes is again the larger weight of high energies which
do have a larger shift but also larger errors. As far as total inverse
$\beta$-decay rates are concerned, the agreement with the results of
Ref.~\cite{Mueller:2011nm} is excellent and no difference reaches more
than $1\,\sigma$.
\begin{table}
\begin{minipage}{\textwidth}
\begin{tabular}{|c|r|rrrr|r r|}
\hline
$^{235}$U&shift&\multicolumn{4}{|c}{$1\,\sigma$ errors [\%]}&\multicolumn{2}{|c|}{$\sigma$ significance}\\
&\%& total &stat&theo.&bias&rate&shape\\
\hline
$\phi_\mathrm{ILL}$&2.4&0.7&0.1&0.6&0.1&3.7&4.5\\
$R_\mathrm{ILL}$&3.7&1.5&$0.0$&1.5&0.1&2.4&3.7\\
\hline
$\phi_\mathrm{MLF}$&0.4&0.7&0.1&0.6&0.1&0.6&6.0\\
$R_\mathrm{MLF}$&1.2&1.5&$0.0$&1.5&0.1&0.8&5.4\\
\hline
\end{tabular}

\begin{tabular}{|c|r|rrrr|r r|}
\hline
$^{239}$Pu&shift&\multicolumn{4}{|c}{$1\,\sigma$ errors [\%]}&\multicolumn{2}{|c|}{$\sigma$ significance}\\
&\%& total &stat&theo.&bias&rate&shape\\
\hline
$\phi_\mathrm{ILL}$&2.9&0.7&0.3&0.6&0.2&3.9&3.2\\
$R_\mathrm{ILL}$&4.2&1.5&0.4&1.4&0.2&2.8&3.2\\
\hline
$\phi_\mathrm{MLF}$&-0.3\footnote{This difference and its sign are driven by the first bin
  from $1.875-2.125$\,MeV and thus the inverse $\beta$-decay rates do
  not reflect this.}&0.7&0.3&-0.6&-0.2&0.5&4.7\\
$R_\mathrm{MLF}$&1.4&1.5&0.4&1.4&0.2&1.0&4.6\\
\hline
\end{tabular}

\begin{tabular}{|c|r|rrrr|r r|}
\hline
$^{241}$Pu&shift&\multicolumn{4}{|c}{$1\,\sigma$ errors [\%]}&\multicolumn{2}{|c|}{$\sigma$ significance}\\
&\%& total &stat&theo.&bias&rate&shape\\
\hline
$\phi_\mathrm{ILL}$&3.2&0.8&0.3&0.7&0.2&4.0&2.6\\
$R_\mathrm{ILL}$&4.7&1.6&0.2&1.5&0.2&3.0&2.6\\
\hline
$\phi_\mathrm{MLF}$&0.4&0.8&0.3&0.7&0.2&0.5&3.7\\
$R_\mathrm{MLF}$&1.0&1.6&0.2&1.5&0.2&0.7&3.7\\
\hline
\end{tabular}

\caption{\label{tab:comparison} These tables compare our result in
  relation to the fluxes from
  Refs.~\cite{Schreckenbach:1985ep,VonFeilitzsch:1982jw,Hahn:1989zr}
  labeled as $\phi_\mathrm{ILL}$ or the inverse $\beta$-decay event
  rates $R_\mathrm{ILL}$. The fluxes and event rates from
  Ref.~\cite{Mueller:2011nm} are referred to as $\phi_\mathrm{MLF}$
  and $R_\mathrm{MLF}$. For the column labeled rate we do not include
  any normalization errors since they are common to the underlying
  data, for the column labeled shape we assume a free normalization.
  The significance in the shape column for event rates assumes $10^6$
  inverse $\beta$-decay events.}

\end{minipage}
\end{table}
In the last column of table~\ref{tab:comparison} we provide the ``shape
significance'', {\it i.e.} the square root of the $\chi^2$-difference
between the various flux and inverse $\beta$-decay rates. For this
calculation we assume a free normalization, which is equivalent to
subtracting any rate differences. For the shape difference of fluxes
we assume infinite statistics, whereas for the inverse $\beta$-decay
shape difference we assume $10^6$ events. We see that shape
differences can be large and quite significant at even more than
$5\,\sigma$. This is, however, based only on the flux errors and does
not include further complications like experimental errors or errors
on the isotope composition of the reactor and burn-up. Thus, at this
level of analysis it remains unclear whether an actual neutrino
measurement could distinguish between the various flux models.

Apart from the quantitative difference we discussed previously, there
are important qualitative differences of our result with respect to
earlier inversions. We  present, for the first time, a
careful and detailed error analysis of both the underlying theory of
$\beta$-decays as well as of the inversion procedure itself. We found
that most corrections to the statistical $\beta$-decay shape, like
finite size effects, are well understood and the associated errors are
only a fraction of the correction itself, see section~\ref{sec:wm}. However,
we also found that the treatment of induced currents in $\beta$-decay is
subject to considerable theoretical uncertainty. We studied in detail
the example of weak magnetism as the largest contributor and found by
comparison with data, the impulse approximation, which is the basis for
assuming that the weak magnetism correction is universal for all beta
decays, is apparently only accurate within about 100\% for most nuclei
and fails completely for a small subset of large-$\log ft$ nuclei. For
the analysis presented so far we assumed a 100\% error on the weak
magnetism corrections, {\it i.e.} we neglect the large-$\log ft$
nuclei. If we assume that our limited sample of nuclei were
representative for fission fragments this error would be 10 times
larger and the sole dominating contribution to the overall error
budget, in which case all flux models would be equivalent within these
much larger error bars. Even if we assume further, that the error bars
would remain small, merely adjusting the central value within this
larger range to about $4$ times its current value would allow to
largely eliminate the upward shift and to bring both the calculation
presented in Ref.~\cite{Mueller:2011nm} and our result into
agreement with the previous ILL fluxes. As a consequence the new found
support for sterile neutrinos as presented in
Refs.~\cite{Mention:2011rk,Kopp:2011qd} would be weakened considerably.

We also investigated in detail how errors propagate through the
inversion procedure and quantified the bias in this method by using
synthetic data sets. It is important to realize that the underlying
mathematical problem is ill-posed and thus we are not dealing with a
simple (or complex) fit of a model to data. Therefore, the use of
synthetic data sets as first shown in Ref.~\cite{Vogel:2007} is
essential to gain insight into error propagation and errors inherent to
the method. The assumption that the statistical errors of the
$\beta$-spectrum would find a one-to-one correspondence in the
resulting neutrino spectra is not supported by our Monte Carlo
simulations. Also, the size of the bias is non-negligible and
would be much larger if we had to use $\beta$-spectra binned in
$250\,\mathrm{keV}$ bins. The combination of bias and statistical inversion
errors from our synthetic data sets yields the total conversion error.
We also would like to emphasize that the $\beta$-residuals obtained
here are significantly smaller than in Ref.~\cite{Mueller:2011nm}
and exceed nowhere the errors of the final result.

In all methods of inversion, input from nuclear databases is required,
since the measured $\beta$-spectrum can be fit by a wide range of
different branches by adjusting the nuclear charge. The problem is,
that nuclear databases tend to be incomplete and may contain
significant errors in level assignments, {\it e.g.} the pandemonium
effect. There are different strategies to deal with these shortcomings
and in Ref.~\cite{Mueller:2011nm} the attempt was made to address
these shortcomings directly at the database level by collecting the
best available experimental data and supplement missing data by
appropriate theoretical models. This resulted in a database which is
about 90\% complete and the remainder was treated by inversion, {\it
  i.e.} with virtual branches. This method, if carefully applied, may
have the potential to ultimately yield the most accurate results.
Here, we followed a different strategy, which tries to minimize the
use of nuclear databases as much as possible. In our inversion
procedure which is solely based on virtual branches, the only place at
which information from nuclear databases enters directly is in the
form of the effective nuclear charge $\bar Z$.  Since $\bar Z$ is an
average quantity, it is relatively straightforward to estimate the
impact of both incompleteness, which we know from comparing fission
yields with the known entries in the database, and of errors in data
itself like the pandemonium effect. It turns out, that $\bar Z$ is
amazingly robust with respect to these issues and never exceeds the
error contributions from nuclear theory. Therefore, we reason that
even improved databases would barely change our result.

In combination all our, quantified, error sources lead to total errors
which are close to the ones found previously~\cite{Mueller:2011nm}. At
high energies our errors are consistently larger, whereas at lower
energies they are either slightly smaller for the case of $^{235}$U or
slightly larger for the 2 Plutonium isotopes, where ``slightly'' means of the order
$0.5\%$.

\section{Summary and Conclusion}
\label{sec:conclusion}

We present a new inversion of the total $\beta$-spectra from fission
fragments obtained at
ILL~\cite{Schreckenbach:1985ep,VonFeilitzsch:1982jw,Hahn:1989zr}
employing only virtual $\beta$-branches. Our analysis is based on the
original data~\cite{Schreckenbach:2011}, which have finer binning,
50\,keV for $^{235}$U and 100\,keV for $^{239}$Pu and $^{241}$Pu,
compared to the published
numbers~\cite{Schreckenbach:1985ep,VonFeilitzsch:1982jw,Hahn:1989zr},
but are otherwise identical. In comparison to
Refs.~\cite{Schreckenbach:1985ep,VonFeilitzsch:1982jw,Hahn:1989zr}
and~\cite{Mueller:2011nm} we treat the theoretical $\beta$-spectrum
shape to a higher level of accuracy by including several sub-leading
corrections, not previously considered in this context. Specifically,
we include a more accurate treatment of finite size effects, screening
and the radiative QED correction for the neutrino, for details see
section~\ref{sec:wm}. All of these corrections to a allowed
$\beta$-spectrum are applied at a branch-by-branch level, which was
not done in
Refs.~\cite{Schreckenbach:1985ep,VonFeilitzsch:1982jw,Hahn:1989zr},
where only an average correction was applied. The only point at which
nuclear structure information enters this calculation is in the form
of the effective nuclear charge $\bar Z$, which is derived in
section~\ref{sec:zeff}. Combining all these ingredients we find the
fluxes as listed in tables~\ref{tab:u235}-\ref{tab:pu241}. Our results
averaged over all energies are in very good agreement with the results
found in Ref.~\cite{Mueller:2011nm} and confirm the overall 2-3\%
shift relative to the original
inversion~\cite{Schreckenbach:1985ep,VonFeilitzsch:1982jw,Hahn:1989zr},
while at the same time the $\beta$-spectrum residuals are very small,
which was not the case in previous analyzes~\cite{Mueller:2011nm}.

We put a particular emphasis on evaluating the errors of the final
neutrino fluxes. We analyzed the associated theory errors and found,
that only the contributions from induced currents, for which we
studied the weak magnetism term as the leading contribution, have
appreciable theory errors. Our overall theory error is similar to the
values previously
quoted~\cite{Schreckenbach:1985ep,VonFeilitzsch:1982jw,Hahn:1989zr,Mueller:2011nm},
however the origin is different, in our case, it being solely due to
the difficulties to estimate the size of the weak magnetism
correction. We also point out, that there are certain nuclei like
$^{14}$C in which the weak magnetism correction is anomalously large
due to a suppression of the Gamow-Teller matrix element. A small
abundance of nuclei of this kind in fission fragments could enhance
the weak magnetism term greatly, see section~\ref{sec:wm}. As a matter
of fact, this enhancement could be large enough to account for a large
fraction of the flux shift found here and in
Ref.~\cite{Mueller:2011nm} and thus may provide a Standard Model
explanation of the reactor anomaly~\cite{Mention:2011rk} without the
need for sterile neutrinos.  While we are not advocating this solution
to the reactor anomaly throughout this work, it seems that Occam's
razor warrants a closer look into this possibility. We provide an
estimate of the effects incompleteness and inevitable inaccuracies of
the nuclear structure data have on the determination of the effective
nuclear charge $\bar Z$, see section~\ref{sec:zeff}.  We evaluate the
inversion errors by using synthetic data sets based on cumulative
fission yields and the ENSDF database and use Monte Carlo simulations
to ``measure'' the bias and statistical error. We find, that the bias
generated by our method is quite small, in agreement with earlier
results~\cite{Vogel:2007}. The statistical errors are sizable and
typically several times larger in the neutrino spectrum than in the
underlying $\beta$-spectrum, see section~\ref{sec:bias}. The resulting
total errors are similar to previous ones~\cite{Mueller:2011nm} at low
energies, however they are larger for the few highest energy bins.
Based on these errors we find averaged over all energies excellent
agreement with the results of Ref.~\cite{Mueller:2011nm}, however if
we look at the energy dependence we do find significant shape
differences, particular at high energies. These difference should be
accessible to measurement, provided a large enough neutrino event
sample of the order $10^6$ events can be obtained, see
table~\ref{tab:comparison}.

As far as the origin of the shift relative to the original
results~\cite{Schreckenbach:1985ep,VonFeilitzsch:1982jw,Hahn:1989zr}
is concerned, we find that both the form of the effective nuclear
charge $\bar Z$ and the use of an average correction to the allowed
$\beta$-shape contribute at a similar level. There are now two
independent and complementary analyzes, which find the same sign and
approximate magnitude of this shift.  However, relevant differences
between the two analyzes do exist, which can be briefly summarized as:
our analysis uses finer binned input data and consequently has much
smaller $\beta$-spectrum residuals; we use a more detailed treatment
of corrections to the allowed $\beta$-spectrum shape and derive the
associated errors from this more detailed treatment; we compute the
inversion errors, both bias and statistical, based on synthetic data
sets; finally, we estimate the errors due to incomplete or incorrect
nuclear structure data.

The aforementioned differences make it seem worthwhile to analyze
existing reactor data using our results in a spirit similar to
Ref.~\cite{Mention:2011rk} and to investigate whether an actual
neutrino spectrum measurement can distinguish between the two
calculations~\cite{Huber:2011xx}. The differences tend to be largest
at high neutrino energies, where typically also the largest
differences between the neutrino fluxes from the various isotopes are
found~\cite{Huber:2004xh}, thus the impact of these new fluxes and
associated errors on neutrino safeguards schemes warrants further
study.

\acknowledgments

Foremost, I would like to thank K.~Schreckenbach for his generosity to
make his original data available to me and all the information he
provided along with it. I also would like to thank D.~Lhuillier and
collaborators for their extensive, helpful comments and for making the
$\beta$-endpoints, branching fractions and simple $\beta$-spectra
derived from the ENSDF database available. I need to thank B.~Holstein
for his clarifications on weak magnetism and induced currents.
Furthermore, I would like to acknowledge useful discussions and
exchanges with J.~Link, L.~Piilonen, R.~Raghavan. I am particularly
indebted to P. Vogel who pointed out significant errors in
Tab.~\ref{tab:wm} in an earlier version of this manuscript. This work
has been supported by the U.S.  Department of Energy under award
number \protect{DE-SC0003915}.

%%%%%%%%%%%%%%%%%%%%%%%%%%%%%%%%%%%%%%%%%%%%%%%%%%%%%%%%%%%%%%%%%%%
\bibliographystyle{apsrev} 
\bibliography{references}
%%%%%%%%%%%%%%%%%%%%%%%%%%%%%%%%%%%%%%%%%%%%%%%%%%%%%%%%%%%%%%%%%%%

\appendix

\section{Supporting Data}
\label{sec:sup}

\begin{table}[h]
\begin{tabular}{|c|rrr rrr|}
\hline
&$b_1$&$b_2$&$b_3$&$b_4$&$b_5$&$b_6$\\
\hline
$a_{-1}$& 0.115 & -1.8123 & 8.2498 & -11.223 & -14.854 & 32.086 \\
$a_{0}$& -0.00062 & 0.007165 & 0.01841 & -0.53736 & 1.2691 & -1.5467 \\
$a_{1}$& 0.02482 & -0.5975 & 4.84199 & -15.3374 & 23.9774 & -12.6534 \\
$a_{2}$& -0.14038 & 3.64953 & -38.8143 & 172.137 & -346.708 & 288.787 \\
$a_{3}$& 0.008152 & -1.15664 & 49.9663 & -273.711 & 657.629 & -603.703\\
$a_{4}$& 1.2145 & -23.9931 & 149.972 & -471.299 & 662.191 & -305.68 \\
$a_{5}$& -1.5632 & 33.4192 & -255.133 & 938.53 & -1641.28 & 1095.36\\
\hline
\end{tabular}
\caption{\label{tab:l0} A reproduction of table~1 of Ref.~\cite{Wilkinson:1990}.}
\end{table}

\begin{table}[h]
\begin{tabular}{|c|ccccc ccccc|}
\hline
 $\tilde Z$&1 & 8 & 13 & 16 & 23 & 27 & 29 & 49 & 84 & 92 \\
 $N(\tilde Z)$&1.000 & 1.420 & 1.484 & 1.497 & 1.52 & 1.544 & 1.561 & 1.637 & 1.838& 1.907\\
\hline
\end{tabular}
\caption{\label{tab:screening} A reproduction of table~4.7 of Ref.~\cite{Behrens:1982}.}
\end{table}

\newpage
\section{The Fluxes}
\label{sec:fluxes}

\begin{table}[h]
\begin{tabular}{|c|c| c|c| c| c| c| c| c| c|}
\hline
\multicolumn{4}{|c|}{value}&\multicolumn{6}{c|}{$1\,\sigma$ errors}\\
\hline
$E_\nu$&$\beta$-res.&$N_{\bar\nu}$ bias&$N_{\bar\nu}$&stat.&bias err.&$\bar Z$&WM&norm.&total\\
MeV&\%&\%&$\mathrm{fission}^{-1}\,\mathrm{MeV}^{-1}$&\%&\%&\%&\%&\%&\%\\
\hline
$2.$&$0$&$-0.3$&$1.32$&$0.57$&$0.055$&$~_{-0.008}^{+0}$&$~_{-0.2}^{+0.14}$&$1.7$&$1.8$\\[0.5ex]
$2.25$&$-0.5$&$-0.3$&$1.12$&$0.63$&$0.18$&$~_{0}^{+0.0074}$&$~_{-0.063}^{+0.017}$&$1.7$&$1.8$\\[0.5ex]
$2.5$&$-0.6$&$0.1$&$9.15\times 10^{\text{-1}}$&$0.66$&$0.21$&$~_{0}^{+0.012}$&$~_{-0.11}^{+0.076}$&$1.7$&$1.9$\\[0.5ex]
$2.75$&$-0.4$&$-0.1$&$7.7\times 10^{\text{-1}}$&$0.67$&$0.037$&$~_{0}^{+0.0047}$&$~_{-0.24}^{+0.21}$&$1.7$&$1.9$\\[0.5ex]
$3.$&$0.1$&$0.2$&$6.51\times 10^{\text{-1}}$&$0.71$&$0.13$&$~_{-0.013}^{+0}$&$~_{-0.36}^{+0.35}$&$1.8$&$1.9$\\[0.5ex]
$3.25$&$0$&$-0.1$&$5.53\times 10^{\text{-1}}$&$0.7$&$0.073$&$~_{-0.043}^{+0}$&$0.49$&$1.8$&$2.$\\[0.5ex]
$3.5$&$0$&$-0.2$&$4.54\times 10^{\text{-1}}$&$0.74$&$0.09$&$~_{-0.083}^{+0}$&$~_{-0.62}^{+0.63}$&$1.8$&$2.$\\[0.5ex]
$3.75$&$-0.3$&$0.4$&$3.64\times 10^{\text{-1}}$&$0.76$&$0.13$&$~_{-0.13}^{+0}$&$~_{-0.75}^{+0.77}$&$1.8$&$2.1$\\[0.5ex]
$4.$&$-0.5$&$0.3$&$2.94\times 10^{\text{-1}}$&$0.77$&$0.061$&$~_{-0.2}^{+0}$&$~_{-0.87}^{+0.91}$&$1.8$&$2.1$\\[0.5ex]
$4.25$&$-0.1$&$-0.3$&$2.3\times 10^{\text{-1}}$&$0.93$&$0.081$&$~_{-0.27}^{+0}$&$1.$&$1.8$&$2.3$\\[0.5ex]
$4.5$&$0.3$&$0$&$1.79\times 10^{\text{-1}}$&$1.2$&$0.11$&$~_{-0.36}^{+0}$&$~_{-1.1}^{+1.2}$&$1.8$&$2.5$\\[0.5ex]
$4.75$&$-0.3$&$0.4$&$1.38\times 10^{\text{-1}}$&$1.2$&$0.12$&$~_{-0.45}^{+0}$&$1.3$&$1.8$&$2.5$\\[0.5ex]
$5.$&$-0.1$&$0.4$&$1.1\times 10^{\text{-1}}$&$1.1$&$0.047$&$~_{-0.56}^{+0}$&$~_{-1.4}^{+1.5}$&$1.8$&$2.6$\\[0.5ex]
$5.25$&$0.2$&$0.4$&$8.64\times 10^{\text{-2}}$&$0.98$&$0.039$&$~_{-0.68}^{+0}$&$~_{-1.5}^{+1.6}$&$1.8$&$2.6$\\[0.5ex]
$5.5$&$-0.1$&$1.$&$6.46\times 10^{\text{-2}}$&$1.1$&$0.087$&$~_{-0.81}^{+0}$&$~_{-1.6}^{+1.7}$&$1.8$&$~_{-2.8}^{+2.7}$\\[0.5ex]
$5.75$&$0$&$0.7$&$5.1\times 10^{\text{-2}}$&$1.2$&$0.081$&$~_{-0.95}^{+0}$&$~_{-1.8}^{+1.9}$&$1.8$&$~_{-3.}^{+2.9}$\\[0.5ex]
$6.$&$0.4$&$0.3$&$3.89\times 10^{\text{-2}}$&$1.5$&$0.041$&$~_{-1.1}^{+0}$&$~_{-1.9}^{+2.}$&$1.8$&$~_{-3.2}^{+3.1}$\\[0.5ex]
$6.25$&$-0.2$&$-0.3$&$2.87\times 10^{\text{-2}}$&$1.6$&$0.14$&$~_{-1.3}^{+0}$&$~_{-2.}^{+2.2}$&$1.9$&$~_{-3.4}^{+3.3}$\\[0.5ex]
$6.5$&$-0.3$&$-0.4$&$2.17\times 10^{\text{-2}}$&$1.4$&$0.11$&$~_{-1.4}^{+0}$&$~_{-2.1}^{+2.3}$&$1.9$&$~_{-3.5}^{+3.3}$\\[0.5ex]
$6.75$&$-0.3$&$-0.1$&$1.61\times 10^{\text{-2}}$&$1.5$&$0.078$&$~_{-1.6}^{+0}$&$~_{-2.3}^{+2.4}$&$1.9$&$~_{-3.7}^{+3.4}$\\[0.5ex]
$7.$&$0.1$&$0.1$&$1.14\times 10^{\text{-2}}$&$1.7$&$0.22$&$~_{-1.8}^{+0}$&$~_{-2.4}^{+2.6}$&$1.9$&$~_{-3.9}^{+3.6}$\\[0.5ex]
$7.25$&$-0.1$&$0.1$&$7.17\times 10^{\text{-3}}$&$2.4$&$0.24$&$~_{-2.}^{+0}$&$~_{-2.5}^{+2.7}$&$1.9$&$~_{-4.5}^{+4.1}$\\[0.5ex]
$7.5$&$0.1$&$-0.6$&$4.64\times 10^{\text{-3}}$&$2.6$&$0.39$&$~_{-2.2}^{+0}$&$~_{-2.7}^{+2.8}$&$1.9$&$~_{-4.8}^{+4.3}$\\[0.5ex]
$7.75$&$0.2$&$-0.7$&$2.97\times 10^{\text{-3}}$&$3.1$&$0.19$&$~_{-2.5}^{+0}$&$~_{-2.8}^{+3.}$&$1.9$&$~_{-5.2}^{+4.7}$\\[0.5ex]
$8.$&$0.1$&$-2.5$&$1.62\times 10^{\text{-3}}$&$5.7$&$0.55$&$~_{-2.7}^{+0}$&$~_{-2.9}^{+3.1}$&$1.9$&$~_{-7.2}^{+6.8}$\\[0.5ex]
\hline
\end{tabular}
\caption{\label{tab:u235} Results for the $^{235}$U anti-neutrino spectrum. This spectrum corresponds to 12\,h irradiation time. The $\beta$-residuals are given for information only and they do not enter into the computation of the errors. The errors in columns~5 and~6 are fully uncorrelated, whereas the errors in columns~7 and~8 are fully correlated between bins and isotopes. The errors in column~9 are fully correlated between bins and isotopes.}
\end{table}

\begin{table}
\begin{tabular}{|c|c| c|c| c| c| c| c| c| c|}
\hline
\multicolumn{4}{|c|}{value}&\multicolumn{6}{c|}{$1\,\sigma$ errors}\\
\hline
$E_\nu$&$\beta$-res.&$N_{\bar\nu}$ bias&$N_{\bar\nu}$&stat.&bias err.&$\bar Z$&WM&norm.&total\\
MeV&\%&\%&$\mathrm{fission}^{-1}\,\mathrm{MeV}^{-1}$&\%&\%&\%&\%&\%&\%\\
\hline
$2.$&$-0.6$&$0$&$1.08$&$1.7$&$0.24$&$~_{0}^{+0.055}$&$~_{-0.11}^{+0.073}$&$1.9$&$2.6$\\[0.5ex]
$2.25$&$-0.5$&$0.2$&$9.2\times 10^{\text{-1}}$&$1.7$&$0.43$&$~_{0}^{+0.029}$&$~_{-0.055}^{+0.031}$&$2.$&$2.6$\\[0.5ex]
$2.5$&$0$&$-0.1$&$7.19\times 10^{\text{-1}}$&$1.5$&$0.27$&$~_{-0.000029}^{+0}$&$~_{-0.18}^{+0.17}$&$2.$&$2.5$\\[0.5ex]
$2.75$&$0.2$&$-0.2$&$6.2\times 10^{\text{-1}}$&$1.6$&$0.12$&$~_{-0.032}^{+0}$&$0.31$&$2.$&$2.6$\\[0.5ex]
$3.$&$0.2$&$0$&$5.15\times 10^{\text{-1}}$&$2.$&$0.2$&$~_{-0.068}^{+0}$&$0.44$&$2.1$&$2.9$\\[0.5ex]
$3.25$&$0.2$&$-0.1$&$3.98\times 10^{\text{-1}}$&$2.2$&$0.12$&$~_{-0.11}^{+0}$&$~_{-0.56}^{+0.58}$&$2.1$&$3.1$\\[0.5ex]
$3.5$&$-0.1$&$-0.4$&$3.29\times 10^{\text{-1}}$&$2.4$&$0.0057$&$~_{-0.15}^{+0}$&$~_{-0.69}^{+0.72}$&$2.1$&$3.3$\\[0.5ex]
$3.75$&$0.5$&$0.1$&$2.61\times 10^{\text{-1}}$&$2.3$&$0.2$&$~_{-0.19}^{+0}$&$~_{-0.82}^{+0.86}$&$2.2$&$3.3$\\[0.5ex]
$4.$&$0.6$&$-0.1$&$1.95\times 10^{\text{-1}}$&$2.4$&$0.2$&$~_{-0.24}^{+0}$&$~_{-0.95}^{+0.99}$&$2.2$&$3.4$\\[0.5ex]
$4.25$&$1.2$&$-0.5$&$1.57\times 10^{\text{-1}}$&$3.1$&$0.21$&$~_{-0.29}^{+0}$&$1.1$&$2.2$&$4.$\\[0.5ex]
$4.5$&$1.1$&$-0.5$&$1.13\times 10^{\text{-1}}$&$4.1$&$0.33$&$~_{-0.35}^{+0}$&$~_{-1.2}^{+1.3}$&$2.3$&$4.9$\\[0.5ex]
$4.75$&$1.3$&$0.3$&$8.33\times 10^{\text{-2}}$&$4.2$&$0.54$&$~_{-0.4}^{+0}$&$~_{-1.3}^{+1.4}$&$2.3$&$5.$\\[0.5ex]
$5.$&$0.3$&$0.7$&$6.13\times 10^{\text{-2}}$&$3.8$&$0.16$&$~_{-0.46}^{+0}$&$1.5$&$2.3$&$4.7$\\[0.5ex]
$5.25$&$0.6$&$0.8$&$4.83\times 10^{\text{-2}}$&$4.2$&$0.22$&$~_{-0.53}^{+0}$&$~_{-1.6}^{+1.7}$&$2.4$&$5.1$\\[0.5ex]
$5.5$&$-0.7$&$1.3$&$3.54\times 10^{\text{-2}}$&$4.9$&$0.004$&$~_{-0.6}^{+0}$&$~_{-1.7}^{+1.8}$&$2.4$&$5.7$\\[0.5ex]
$5.75$&$0.1$&$0.8$&$2.92\times 10^{\text{-2}}$&$5.5$&$0.16$&$~_{-0.67}^{+0}$&$~_{-1.8}^{+2.}$&$2.5$&$6.4$\\[0.5ex]
$6.$&$-0.4$&$0.1$&$1.92\times 10^{\text{-2}}$&$7.8$&$0.28$&$~_{-0.74}^{+0}$&$~_{-2.}^{+2.1}$&$2.5$&$8.5$\\[0.5ex]
$6.25$&$0.2$&$-0.7$&$1.28\times 10^{\text{-2}}$&$8.7$&$0.15$&$~_{-0.82}^{+0}$&$~_{-2.1}^{+2.2}$&$2.5$&$9.4$\\[0.5ex]
$6.5$&$-0.9$&$-0.2$&$9.98\times 10^{\text{-3}}$&$9.1$&$0.39$&$~_{-0.9}^{+0}$&$~_{-2.2}^{+2.4}$&$2.6$&$~_{-9.8}^{+9.7}$\\[0.5ex]
$6.75$&$2.9$&$-0.2$&$7.54\times 10^{\text{-3}}$&$9.9$&$0.12$&$~_{-0.98}^{+0}$&$~_{-2.3}^{+2.5}$&$2.6$&$11.$\\[0.5ex]
$7.$&$-4.$&$0.3$&$4.98\times 10^{\text{-3}}$&$12.$&$0.33$&$~_{-1.1}^{+0}$&$~_{-2.5}^{+2.6}$&$2.6$&$13.$\\[0.5ex]
$7.25$&$-0.6$&$-0.1$&$3.26\times 10^{\text{-3}}$&$18.$&$1.1$&$~_{-1.2}^{+0}$&$~_{-2.6}^{+2.8}$&$2.7$&$18.$\\[0.5ex]
$7.5$&$-0.4$&$-2.3$&$1.95\times 10^{\text{-3}}$&$22.$&$0.78$&$~_{-1.2}^{+0}$&$~_{-2.7}^{+2.9}$&$2.7$&$23.$\\[0.5ex]
$7.75$&$-0.2$&$2.$&$8.47\times 10^{\text{-4}}$&$26.$&$3.1$&$~_{-1.3}^{+0}$&$~_{-2.9}^{+3.1}$&$2.7$&$27.$\\[0.5ex]
$8.$&$-3.1$&$-4.4$&$5.87\times 10^{\text{-4}}$&$28.$&$7.7$&$~_{-1.4}^{+0}$&$~_{-3.}^{+3.2}$&$2.8$&$29.$\\[0.5ex]
\hline
\hline
\end{tabular}
\caption{\label{tab:pu239} Results for the $^{239}$Pu anti-neutrino spectrum. This spectrum corresponds to 36\,h irradiation time.  The $\beta$-residuals are given for information only and they do not enter into the computation of the errors. The errors in columns~5 and~6 are fully uncorrelated, whereas the errors in columns~7 and~8 are fully correlated between bins and isotopes. The errors in column~9 are fully correlated between bins and isotopes.}
\end{table}

\begin{table}
\begin{tabular}{|c|c| c|c| c| c| c| c| c| c|}
\hline
\multicolumn{4}{|c|}{value}&\multicolumn{6}{c|}{$1\,\sigma$ errors}\\
\hline
$E_\nu$&$\beta$-res.&$N_{\bar\nu}$ bias&$N_{\bar\nu}$&stat.&bias err.&$\bar Z$&WM&norm.&total\\
MeV&\%&\%&$\mathrm{fission}^{-1}\,\mathrm{MeV}^{-1}$&\%&\%&\%&\%&\%&\%\\
\hline
$2$&$0$&$0.1$&$1.26$&$1.7$&$0.18$&$~_{0}^{+0.16}$&$~_{-0.18}^{0}$&$1.8$&$2.5$\\[0.5ex]
$2.25$&$0$&$-0.2$&$1.08$&$1.6$&$0.34$&$~_{0}^{+0.13}$&$~_{-0.12}^{0}$&$1.8$&$2.4$\\[0.5ex]
$2.5$&$0.1$&$0.1$&$8.94\times 10^{\text{-1}}$&$1.4$&$0.33$&$~_{0}^{+0.079}$&$~_{-0.23}^{+0.099}$&$1.8$&$2.3$\\[0.5ex]
$2.75$&$0$&$-0.3$&$7.77\times 10^{\text{-1}}$&$1.4$&$0.2$&$~_{0}^{+0.024}$&$~_{-0.34}^{+0.24}$&$1.8$&$2.3$\\[0.5ex]
$3$&$0$&$-0.2$&$6.41\times 10^{\text{-1}}$&$1.6$&$0.12$&$~_{-0.041}^{+0}$&$~_{-0.46}^{+0.38}$&$1.8$&$2.4$\\[0.5ex]
$3.25$&$0$&$-0.3$&$5.36\times 10^{\text{-1}}$&$1.6$&$0.15$&$~_{-0.12}^{+0}$&$~_{-0.57}^{+0.52}$&$1.8$&$2.5$\\[0.5ex]
$3.5$&$0$&$-0.1$&$4.39\times 10^{\text{-1}}$&$1.6$&$0.22$&$~_{-0.2}^{+0}$&$~_{-0.68}^{+0.65}$&$1.8$&$2.5$\\[0.5ex]
$3.75$&$0$&$0.1$&$3.46\times 10^{\text{-1}}$&$1.4$&$0.2$&$~_{-0.29}^{+0}$&$~_{-0.8}^{+0.79}$&$1.8$&$~_{-2.5}^{+2.4}$\\[0.5ex]
$4$&$0$&$0.2$&$2.82\times 10^{\text{-1}}$&$1.5$&$0.11$&$~_{-0.39}^{+0}$&$~_{-0.91}^{+0.93}$&$1.8$&$2.6$\\[0.5ex]
$4.25$&$0$&$-0.3$&$2.2\times 10^{\text{-1}}$&$1.9$&$0.17$&$~_{-0.5}^{+0}$&$~_{-1.}^{+1.1}$&$1.9$&$2.9$\\[0.5ex]
$4.5$&$0.1$&$0$&$1.66\times 10^{\text{-1}}$&$2.5$&$0.29$&$~_{-0.62}^{+0}$&$~_{-1.1}^{+1.2}$&$1.9$&$~_{-3.4}^{+3.3}$\\[0.5ex]
$4.75$&$0$&$0.9$&$1.25\times 10^{\text{-1}}$&$2.5$&$0.48$&$~_{-0.75}^{+0}$&$~_{-1.2}^{+1.3}$&$1.9$&$3.5$\\[0.5ex]
$5$&$0.1$&$1.$&$9.74\times 10^{\text{-2}}$&$2.3$&$0.033$&$~_{-0.89}^{+0}$&$~_{-1.4}^{+1.5}$&$1.9$&$~_{-3.4}^{+3.3}$\\[0.5ex]
$5.25$&$0$&$1.$&$7.47\times 10^{\text{-2}}$&$2.3$&$0.14$&$~_{-1.}^{+0}$&$~_{-1.5}^{+1.6}$&$1.9$&$~_{-3.5}^{+3.4}$\\[0.5ex]
$5.5$&$0$&$1.6$&$5.58\times 10^{\text{-2}}$&$2.6$&$0.13$&$~_{-1.2}^{+0}$&$~_{-1.6}^{+1.8}$&$1.9$&$~_{-3.8}^{+3.6}$\\[0.5ex]
$5.75$&$0$&$0.9$&$4.11\times 10^{\text{-2}}$&$3.2$&$0.21$&$~_{-1.4}^{+0}$&$~_{-1.7}^{+1.9}$&$1.9$&$~_{-4.3}^{+4.2}$\\[0.5ex]
$6$&$-0.2$&$0.2$&$3.05\times 10^{\text{-2}}$&$4.$&$0.09$&$~_{-1.5}^{+0}$&$~_{-1.8}^{+2.}$&$1.9$&$~_{-5.}^{+4.9}$\\[0.5ex]
$6.25$&$0.3$&$-0.6$&$1.98\times 10^{\text{-2}}$&$4.4$&$0.33$&$~_{-1.7}^{+0}$&$~_{-1.9}^{+2.2}$&$1.9$&$~_{-5.5}^{+5.3}$\\[0.5ex]
$6.5$&$-0.1$&$-0.3$&$1.54\times 10^{\text{-2}}$&$4.5$&$0.43$&$~_{-1.9}^{+0}$&$~_{-2.}^{+2.3}$&$1.9$&$~_{-5.7}^{+5.4}$\\[0.5ex]
$6.75$&$0$&$-0.1$&$1.09\times 10^{\text{-2}}$&$4.7$&$0.074$&$~_{-2.1}^{+0}$&$~_{-2.2}^{+2.5}$&$1.9$&$~_{-5.9}^{+5.6}$\\[0.5ex]
$7$&$0$&$0.6$&$7.75\times 10^{\text{-3}}$&$4.8$&$0.39$&$~_{-2.3}^{+0}$&$~_{-2.3}^{+2.6}$&$1.9$&$~_{-6.1}^{+5.8}$\\[0.5ex]
$7.25$&$0.1$&$0.4$&$4.47\times 10^{\text{-3}}$&$6.3$&$0.36$&$~_{-2.5}^{+0}$&$~_{-2.4}^{+2.7}$&$2.$&$~_{-7.5}^{+7.2}$\\[0.5ex]
$7.5$&$-0.3$&$-1.$&$2.9\times 10^{\text{-3}}$&$7.7$&$0.3$&$~_{-2.8}^{+0}$&$~_{-2.5}^{+2.9}$&$2.$&$~_{-8.8}^{+8.4}$\\[0.5ex]
$7.75$&$0$&$-0.4$&$1.78\times 10^{\text{-3}}$&$8.3$&$0.063$&$~_{-3.}^{+0}$&$~_{-2.6}^{+3.}$&$2.$&$~_{-9.4}^{+9.1}$\\[0.5ex]
$8$&$0.7$&$-3.5$&$1.06\times 10^{\text{-3}}$&$12.$&$0.63$&$~_{-3.3}^{+0}$&$~_{-2.7}^{+3.2}$&$2.$&$~_{-13.}^{+12.}$\\[0.5ex]
\hline
\hline
\end{tabular}
\caption{\label{tab:pu241} Results for the $^{241}$Pu anti-neutrino spectrum. This spectrum corresponds to 43\,h irradiation time.  The $\beta$-residuals are given for information only and they do not enter into the computation of the errors. The errors in columns~5 and~6 are fully uncorrelated, whereas the errors in columns~7 and~8 are fully correlated between bins and isotopes. The errors in column~9 are fully correlated between bins and isotopes.}
\end{table}

\end{document}